\documentclass[10pt,journal,compsoc]{IEEEtran}
\IEEEoverridecommandlockouts
\ifCLASSOPTIONcompsoc
  \usepackage[nocompress]{cite}
\else
  \usepackage{cite}
\fi
\usepackage{amsmath,amssymb,amsfonts}
\usepackage{algorithmic}
\usepackage{textcomp}
\usepackage{xcolor}
\usepackage{graphicx}
\usepackage{multirow}
\usepackage{makecell}
\usepackage{csquotes}
\usepackage{framed}
\usepackage{array}
\usepackage[bookmarks=false]{hyperref}
\usepackage[capitalise]{cleveref}
\usepackage{siunitx}
\usepackage{fontawesome}
\usepackage{soul}
\definecolor{greenncs}{rgb}{0.0, 0.62, 0.42}
\definecolor{airforceblue}{rgb}{0.36, 0.54, 0.66}
\colorlet{soulc}{greenncs!30}
\sethlcolor{soulc}
\crefformat{footnote}{#2\footnotemark[#1]#3}

\newcolumntype{L}[1]{>{\raggedright\let\newline\\\arraybackslash\hspace{0pt}}m{#1}}
\newcolumntype{C}[1]{>{\centering\let\newline\\\arraybackslash\hspace{0pt}}m{#1}}
\newcolumntype{R}[1]{>{\raggedleft\let\newline\\\arraybackslash\hspace{0pt}}m{#1}}

\makeatletter
\renewcommand{\@IEEEsectpunct}{\ \,}%
\makeatother

\def\BibTeX{{\rm B\kern-.05em{\sc i\kern-.025em b}\kern-.08em
    T\kern-.1667em\lower.7ex\hbox{E}\kern-.125emX}}
    
\begin{document}

\title{Self-Admitted Technical Debt in the Embedded Systems Industry: An Exploratory Case Study}

\author{Yikun~Li,
        Mohamed~Soliman,
        Paris~Avgeriou,
        and~Lou~Somers
\IEEEcompsocitemizethanks{\IEEEcompsocthanksitem Yikun Li, Mohamed Soliman, Paris Avgeriou are with the Bernoulli Institute for Mathematics, Computer Science and Artificial Intelligence, University of Groningen, The Netherlands.\protect\\
E-mail: \{yikun.li, m.a.m.soliman, p.avgeriou\}@rug.nl
\IEEEcompsocthanksitem Lou~Somers is with the Department of Mathematics and Computer Science, Eindhoven University of Technology, The Netherlands.\protect\\
E-mail: l.j.a.m.somers@tue.nl}
\thanks{Manuscript received; revised. This work was supported by ITEA3 and RVO under grant agreement No. 17038 VISDOM (\url{https://visdom-project.github.io/website}).}}

\IEEEtitleabstractindextext{
\begin{abstract}
Technical debt denotes shortcuts taken during software development, mostly for the sake of expedience.
When such shortcuts are admitted explicitly by developers (e.g., writing a TODO/Fixme comment), they are termed as \emph{Self-Admitted Technical Debt} or \emph{SATD}.
There has been a fair amount of work studying SATD management in Open Source projects, but SATD in industry is relatively unexplored. At the same time, there is no work focusing on developers' perspectives towards SATD and its management.
To address this, we conducted an exploratory case study in cooperation with an industrial partner to study how they think of SATD and how they manage it.
Specifically, we collected data by identifying and characterizing SATD in different sources (issues, source code comments, and commits) and carried out a series of interviews with 12 software practitioners.
The results show: 1) the core characteristics of SATD in industrial projects; 2) developers' attitudes towards identified SATD and statistics; 3) triggers for practitioners to introduce and repay SATD; 4) relations between SATD in different sources; 5) practices used to manage SATD; 6) challenges and tooling ideas for SATD management.
\end{abstract}

}

\maketitle

\IEEEdisplaynontitleabstractindextext

\IEEEpeerreviewmaketitle

\section{Introduction}

Technical debt (TD) refers to compromising the long-term maintainability and evolvability of software systems by selecting sub-optimal solutions, in order to achieve short-term goals  \cite{avgeriou2016managing}.
When software developers incur technical debt, they sometimes explicitly admit it; for example, software developers may write \textit{TODO} or \textit{Fixme} in a source code comment, indicating a sub-optimal solution in that part of the code.
Potdar and Shihab \cite{potdar2014exploratory} called these textual statements \emph{Self-Admitted Technical Debt} (SATD).
SATD can be found in several sources such as source code comments \cite{potdar2014exploratory}, issues in issue tracking systems \cite{dai2017detecting,li2020identification}, and commit messages\cite{zampetti2021self}.

The SATD that can be identified in such sources is complementary to the technical debt that can be identified in source code through static analysis.
This is because, certain types of technical debt cannot be identified by analyzing source code.
For example, \emph{partially implemented requirements} is a type of \emph{requirement debt} \cite{li2020identification} that can be identified from source code comments or issue tracking systems but not from running source code analysis tools:
\textit{``TODO: This class only has partial Undo support (basically just those members that had it as part of a previous implementation) [from Apache ArgoUML\footnote{\url{https://github.com/argouml-tigris-org/argouml/blob/d5cd45cb4409c6f50747a3a2671219111b443c48/src/argouml-app/src/org/argouml/notation/NotationSettings.java\#L108}}].''} Therefore it is important to identify and further manage SATD, in addition to the more traditional approach of managing technical debt in source code.

There has been a fair amount of work investigating the identification \cite{da2017using,ren2019neural}, measurement \cite{kamei2016using}, prioritization \cite{mensah2018value}, and repayment \cite{maldonado2017empirical,zampetti2018self} of SATD.
However, to the best of our knowledge, all previous studies (but one, namely \cite{zampetti2021self}) identified SATD in open-source projects; we actually know little about SATD in industrial projects.
Moreover, none of the previous studies has surveyed software developers about SATD, in order to capture their perspectives towards SATD management, and tooling support for different sources.
Without involving software developers to investigate SATD, researchers risk developing theories or approaches, which do not align with the needs and practices of software engineers.

To address these shortcomings, we conducted an exploratory case study in collaboration with an industrial partner to investigate how SATD is managed and how this can be supported.
We collected data in two steps.
First, we identified and characterized SATD in projects within that company from three sources: issues, source code comments, and commits. This step took place by using pre-trained machine learning models \cite{li2022automatic}.
Second, we carried out a series of interviews with 12 software practitioners from that organization to understand their perception of what SATD really is, how it is managed, and how this management can be potentially improved.
The contributions and main findings of this study are summarized as follows:

\begin{itemize}
    \item \textbf{Characterizing SATD in industrial projects.}
    The results indicate that most technical debt is admitted in issues, followed by source code comments and commit messages.
    Non-SATD issues take a significantly shorter time to close, compared to SATD issues.

    \item \textbf{Reporting developers' attitudes towards identified SATD and statistics.}
    Most interviewees acknowledged the identified SATD.
    However, they do need more information to assess the importance of individual SATD items.

    \item \textbf{Reporting relations between SATD from different sources.}
    We found that SATD in code comments and issues is referenced in the other sources, while SATD in commits is not referenced in other sources.
    
    \item \textbf{Reporting triggers on SATD introduction and repayment.}
    The results show developers have different reasons to introduce and pay back SATD, depending on the data source (code comments, issues, commits).

    \item \textbf{Reporting practices used to manage SATD.}
    We summarize and report practices that are used to assist in SATD prioritization and repayment.

    \item \textbf{Reporting tooling support for SATD management.}
    We report tool features that developers suggested as useful for SATD identification, traceability, prioritization, and repayment.
\end{itemize}

The rest of this paper is organized as follows. 
\cref{sec:related} discusses related work.
The case study design is elaborated in \cref{sec:approach}.
\cref{sec:results} presents the results, and \cref{sec:discussion} discusses the implications of these results on researchers and practitioners.
Finally, threats to validity are evaluated in \cref{sec:validity} and conclusions and future work are drawn in \cref{sec:conclusion}.

\section{Related Work}
\label{sec:related}

To facilitate comparison to our work, we split the related work into two parts: work associated with SATD in Open-Source Software and work associated with SATD in industrial settings.

\subsection{SATD in Open-Source Software}

Potdar and Shihab \cite{potdar2014exploratory} were the first to study SATD in source code comments.
They analyzed four open-source projects and identified SATD in them.
They found that 2.4\% to 31\% of source files contain SATD comments and only 26.3\% to 63.5\% of SATD are removed after introduction.
Moreover, the results of Potdar and Shihab show that experienced developers tend to introduce more SATD compared to inexperienced developers.
Building on this work, Maldonado and Shihab \cite{maldonado2015detecting} focused on the types of SATD in open-source projects.
They analyzed 33K code comments from five projects and categorized SATD into five categories: design, requirement, defect, documentation, and test debt.
The results indicated that design debt is the most common type of SATD, as 42\% to 84\% of classified SATD is design debt.

Subsequently, there was a significant focus on automatic SATD identification.
Maldonado \textit{et al.} \cite{da2017using} manually classified source code comments into different types of SATD from ten open-source projects and utilized the maximum entropy classifier to automatically identify design debt and requirement debt.
Similarly, Huang \textit{et al.} \cite{huang2018identifying} used the feature selection method to select the most important features and adopted the ensemble learning technique to leverage different machine learning approaches to accurately identify SATD, again from source code comments.
Furthermore, different machine learning approaches were applied to achieve higher predictive performance for SATD identification.
Specifically, Ren \textit{et al.} \cite{ren2019neural} proposed a Convolutional Neural Network-based approach to accurately identify SATD from source code comments.
Wang \textit{et al.} \cite{wang2020detecting} proposed an attention-based neural network to automatically detect SATD.
In addition to using source code comments, few studies focused on identifying SATD from other sources.
Dai and Kruchten \cite{dai2017detecting} manually analyzed 8K issue tickets and used the Naive Bayes method to classify SATD issues and non-SATD issues.
In our previous work \cite{li2022identifying}, we examined 23K issue sections and proposed a Convolutional Neural Network-based approach to identify SATD from issue tracking systems.

In addition to SATD identification, there has been work related to the measurement, prioritization, and repayment of SATD.
Kamei \textit{et al.} \cite{kamei2016using} explored ways to measure the interest of SATD and suggested using LOC and Fan-In measures.
The results indicated that 42.2\% to 44.2\% of SATD incurs positive interest (i.e. technical debt costs more to repay in the future), while 8.1\% to 13.8\% of SATD incurs negative interest (i.e., technical debt costs less to pay back in the future).
Mensah \textit{et al.} \cite{mensah2018value} introduced a SATD prioritization scheme which consists of identification, examination, and rework effort estimation.
The results showed that a rework effort of modifying 10 to 25 commented LOC per SATD source file is required for highly prioritized SATD tasks.
Besides, Maldonado \textit{et al.} \cite{maldonado2017empirical} analyzed five open-source projects to investigate the repayment of SATD.
The results indicated that most of SATD is removed eventually and the payback is mostly done by those that incurred the SATD in the first place.
They also found that SATD lingers in the code for approximately 18 to 172 days.
In a similar study, Zampetti \textit{et al.} \cite{zampetti2018self} looked into how SATD is resolved in five open-source projects.
They found that between 20\% to 50\% of SATD comments are removed by accident, and 8\% of SATD repayment is documented in commit messages.

Compared to all aforementioned studies, our study has the following differences: a) we utilized machine learning models to identify and characterize SATD; b) we performed this analysis on a number of  different sources, instead of only one; c) we work in an industrial setting instead of open-source systems; d) we explored developers' perspectives towards both the nature of SATD and its management.

\subsection{SATD in Industrial Settings}

SATD in industrial settings is relatively unexplored; there is only one work that studied SATD in industrial settings and compared it with open-source settings \cite{zampetti2021self}.
Specifically, Zampetti \textit{et al.} surveyed 52 industrial developers and 49 open-source project contributors.
They focused on technical debt admitted in source code comments and found that technical debt annotation practices and the typical content of SATD comments are similar in industrial and open-source settings.
Furthermore, the results showed that admitted technical debt in industrial projects is implicitly discouraged by the fear of taking on responsibilities.
The results indicated that technical debt is also admitted in other sources, including, among others, commit messages, pull requests, and issue trackers.

In contrast to Zampetti \textit{et al.} \cite{zampetti2021self}, who only investigated source code comments, we focus on analyzing SATD from multiple sources (i.e., source code comments, commit messages, and issue tracking systems). 
In addition, we use machine learning techniques to identify SATD from different sources in industrial settings, demonstrate the characteristics of SATD, and present the interviewees' attitudes towards the identified SATD and statistics.
Finally, we also study the process of managing SATD, as well as how to improve it from the point of view of software practitioners.

\section{Study Design}
\label{sec:approach}

\subsection{Objective and Research Questions}
\label{sec:objective}

The goal of this study, formulated according to the Goal-Question-Metric \cite{Solingen:02} template is to ``\textit{\textbf{analyze} self-admitted technical debt in source code comments, issue tracking systems, and commit messages \textbf{for the purpose of} understanding and improvement \textbf{with respect to} the nature and management process of self-admitted technical debt in practice \textbf{from the point of view of} software engineers \textbf{in the context of} the embedded systems industry.}''
To be more precise, we aim at understanding both the nature of SATD per se and the process of managing it, as well as improving this process. Consequently, we formulate three main research questions (RQs) that are further refined into sub-questions.
In \cref{sec:results} we will not answer the main RQs directly, but only indirectly through answering the sub-questions.

\begin{figure*}[thpb]
  \centering
  \includegraphics[width=\linewidth]{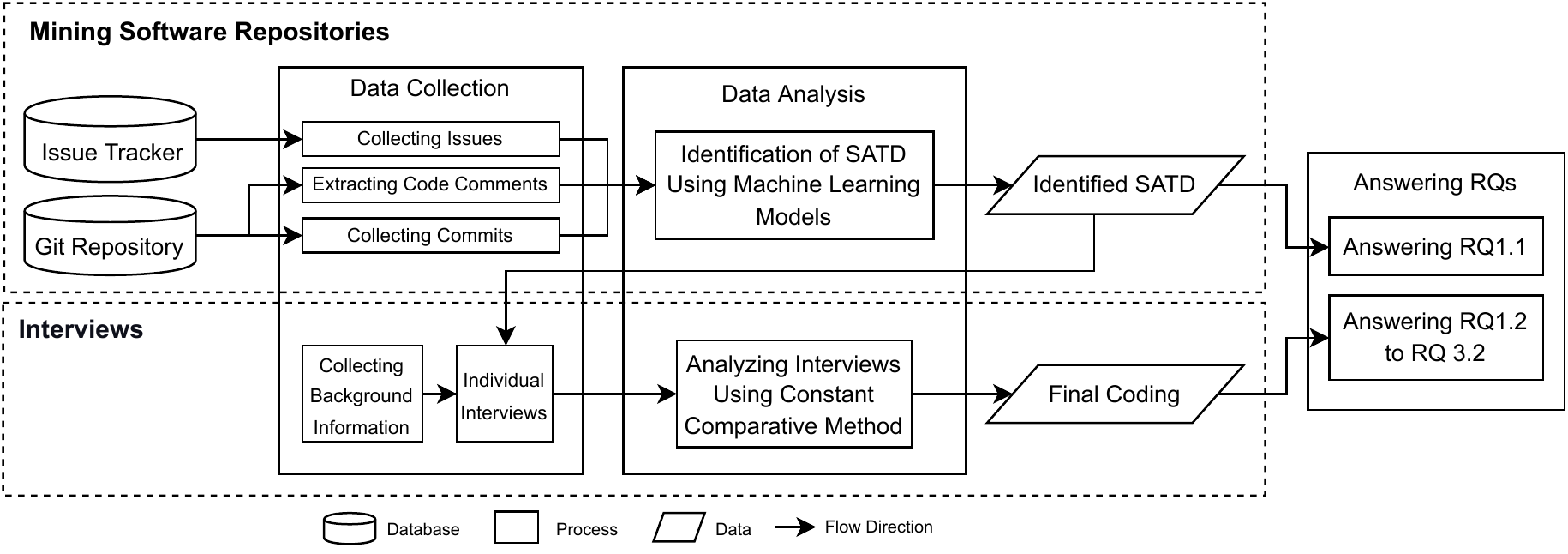}
  \caption{The framework of our approach.}
  \label{fig:framework}
\end{figure*}

\begin{itemize}
    \item \textbf{RQ1: What is the nature of SATD in industry?}

    \begin{itemize}
        \item \textbf{RQ1.1:} 
        \textit{What are the types, amounts, resolution time, and sources of SATD items in industrial settings?}
        \\
        \textbf{Rationale:}
        There are significant differences between open-source projects and industrial projects concerning project management, tooling, as well as collaboration and communication \cite{yamashita2015inter,bosu2017process}.
        Thus, developers may admit technical debt differently in the two cases. 
        Meanwhile, as mentioned in \cref{sec:related}, all (but one, namely \cite{zampetti2021self}) previous studies on SATD have focused on open-source projects.
        Thus, 
        determining the types of SATD (e.g. requirements, design, code debt), amount of SATD (i.e. number and percentages of SATD items), and the sources of SATD (e.g. issue tracker or source code comments)
        in industrial projects, and comparing them with open-source projects could help researchers understand what SATD looks like in practice, and practitioners to better manage SATD in both cases.
    
        \item \textbf{RQ1.2:} \textit{How is automatically identified SATD regarded by professional software engineers?}\\
        \textbf{Rationale:}
        The identification of SATD can be automated, e.g. by using machine learning techniques \cite{ren2019neural,li2022identifying}.
        However, as far as software engineers are concerned, the identified SATD items could be obsolete, inaccurate, irrelevant, or inconsistent with the code.
        We aim at understanding how far software engineers consider that the SATD items are indeed important and relevant for their system.
        We also want to understand whether software engineers agree with the main statistics of the identified SATD (e.g. percentage of backlog items and the lifetime of SATD items). 
        This can help us understand the strengths and weaknesses of automated SATD identification.
        
        \item \textbf{RQ1.3:} \textit{What are the relations between SATD in different sources?}\\
        \textbf{Rationale:}
        The different sources where SATD is documented (e.g. source code, issues, commits), are implicitly or explicitly related to each other.
        Thus, developers sometimes choose to document the same technical debt in more than one source.
        For example, when developers encounter SATD in code comments and they consider this debt as important, they might document it in issue tracking systems for more exposure and visibility.
        In these cases, we are interested in understanding the connections between the SATD items in these different sources.
        This can assist in improving the traceability of SATD in different sources.
    
    \end{itemize}
    
    \item \textbf{RQ2: How is SATD being managed in industry?}
    
    \begin{itemize}
        \item \textbf{RQ2.1:} \textit{When is technical debt (not) admitted in source code comments, issue tracking systems, and commit messages?}\\
        \textbf{Rationale:}
        It is important to understand the reasons for documenting or not documenting technical debt in different sources.
        This can help researchers in coming up with guidelines and practices for SATD documentation.
        Furthermore, this could help develop tools to assist in documenting SATD.
    
        \item \textbf{RQ2.2:} \textit{What are the pros and cons of admitting technical debt in different sources?}\\
        \textbf{Rationale:}
        Each source has its advantages and disadvantages in terms of documenting technical debt.
        For example, technical debt that is admitted in code comments, allows developers reading those comments to also examine the problem in the adjacent code.
        On the downside, SATD in code comments has limited visibility for team leads and project managers and typically does not get added to the backlog.
        Understanding the pros and cons of different sources for documenting SATD could help developers make better use of different sources to document technical debt.
    
        \item \textbf{RQ2.3:} \textit{What are the triggers to pay back or not pay back SATD?}\\
        \textbf{Rationale:}
        Developers are more likely to pay back SATD in certain cases. We plan to investigate the developers' motivation both for repayment and for deciding to leave SATD in the system.
        This could help researchers understand the reasons for SATD repayment and develop a tool to assist practitioners in paying back TD.
    
        \item \textbf{RQ2.4:} \textit{What practices are used to support SATD management in industrial settings?}\\
        \textbf{Rationale:}
        While, a number of studies have investigated TD management in industry, we know very little about managing self-admitted TD.
        This RQ can help in understanding the current practices of SATD management in industrial settings.
        For example, developers could group similar SATD to facilitate the SATD repayment.
        Software practitioners may be able to use some of these practices in their own context, while researchers may investigate ways of supporting them.
    \end{itemize}
    
    \item \textbf{RQ3: How can we improve SATD management?}
    
    \begin{itemize}
        \item \textbf{RQ3.1:} \textit{What challenges do software practitioners face when managing SATD?}\\
        \textbf{Rationale:}
        Understanding the challenges of SATD management can help researchers to provide support for addressing those challenges.
        For example, prioritization of technical debt items is a typical challenge in any kind of technical debt, including self-admitted. If we obtain an in-depth understanding of why it is difficult to prioritize SATD in specific, we are in a better position to propose practices, tools, or guidelines to address this challenge.
    
        \item \textbf{RQ3.2:} \textit{What features should tools have to effectively manage SATD?}\\
        \textbf{Rationale:}
        As mentioned in \cref{sec:related}, there are tools supporting SATD identification.
        However, we currently lack tools to assist in other activities of SATD management such as prioritization or repayment \cite{SIERRA201970}.
        Answering this question can support the development of new tools or the improvement of existing ones that could help practitioners better and easier manage SATD.
    \end{itemize}
\end{itemize}

\cref{fig:framework} presents the overall framework of our approach to answering the research questions.
The two major processes (i.e., data collection and data analysis) are elaborated in the following sub-sections.

\subsection{Cases and Units of Analysis}

This case study is designed as a single embedded case study \cite{runeson2012case}.
Our case is a large software company in the \emph{embedded systems industry} that chooses to remain anonymous. 
The software development in this company adopts \emph{Scrum} development practices.

Because we focus on understanding SATD and its management process, as well as improving the latter, we collected data from two types of units.
The first type of unit is software artifacts, including source code, commits, and issue tracking systems.
It is noted that the studied projects mainly use C++ and XML files.
There are, on average, 20 software engineers working on the analyzed projects.
We identified and analyzed the nature of SATD from these software artifacts.
The second type of unit is software engineers that participated in the development of a specific project.
More specifically, each engineer represents a single unit.
We conducted interviews with software engineers to derive their opinions on the aforementioned SATD nature, as well as to understand and improve SATD management.
Details about the background of the practitioners are presented in \cref{tb:backgroud}.

\begin{table*}[th]
\caption{Questions for Individual Interviews.}
\label{tb:questions}
\begin{center}
\resizebox{2\columnwidth}{!}{
\def\arraystretch{1.2}
\begin{tabular}{@{\extracolsep{4pt}}L{13.2cm}C{1.8cm}@{}}
\hline
\textbf{Question} & \textbf{Related RQs} \\
\hline
Do you acknowledge the identified SATD items from the different sources? & RQ1.2 \\
Do you consider identified SATD items to be important? & RQ1.2 \\
What do you think of different types of SATD? & RQ1.2 \\
What do you think about the average time to close different types of SATD issues and non-SATD issues? & RQ1.2 \\
Do these SATD items and statistics give you new information or insights about this project? & RQ1.2 \\
\hline
What do you think are the relations between TD documented in different sources? & RQ1.3 \\
\hline
Do you record TD in your project? & RQ2.1, RQ2.2 \\
Which TD items do you usually record and which do not? & RQ2.1, RQ2.2 \\
Where do you typically record TD? & RQ2.1, RQ2.2 \\
Do you have any constraints on recording TD? & RQ2.1, RQ2.2 \\
Which types of TD do you usually record? & RQ2.1, RQ2.2 \\
What do you think are the differences between TD documented in different sources? & RQ2.1, RQ2.2 \\
\hline
How do you decide on resolving one of the recorded TD items? & RQ2.3 \\
\hline
How do you manage the recorded TD items in practice? & RQ2.4 \\
How do you prioritize documented TD items in practice? & RQ2.4 \\
What strategies do you follow to pay back documented TD? & RQ2.4 \\
\hline
What challenges do you encounter when you manage recorded TD? & RQ3.1 \\
\hline
What features would you like to have from an ideal tool to manage SATD? & RQ3.2 \\
\hline
\end{tabular}
}
\end{center}
\vspace{-3mm}
\end{table*}

\subsection{Data Collection}

As seen in \cref{fig:framework}, data is collected through \textit{analysis of work artifacts} and \textit{interviews}, which are third- and first degree data collection methods respectively, according to Lethbridge \textit{et al.} \cite{Lethbridge2005studying}.
These two methods are explained in the following subsections in detail.

\subsubsection{Analysis of Work Artifacts}

In order to answer the research questions, we choose a large-scale industrial project which contains eight sub-projects.
More specifically, the selected project has over 475K lines of comments, 21K commits, and 130K files (including documentation, test files, configuration files, etc.).
Regarding issues, we collected 78K issues from the issue tracking system.
We note that, all embedded software in the selected company uses the same issue tracking system; thus, the collected issues come from all embedded software while comments and commits are from the selected project where we had access.
Because we focus on three different types of work artifacts, namely source code comments, issue tracking systems, and commits, we obtained data from these different sources separately.
For source code comments, we created a script to first retrieve all code changes in git and then extract all source code comments using the \emph{CommentParser} tool\footnote{\url{https://github.com/jeanralphaviles/comment_parser}}.
We manually verified the correctness of the extracted comments with this tool before collecting the data.
For issue tracking systems, we extracted all issue descriptions for analysis using the API of the issue tracker used by the company (Microsoft TFS).
Lastly, for commits, we obtained all commit descriptions in git.
The scripts for collecting data are included in the replication package\footnote{\label{l:data}\url{https://github.com/yikun-li/satd-in-industry}}.
We analyzed the latest version of the selected industrial project on July 7th, 2021.

\subsubsection{Interviews}

We have conducted semi-structured interviews to collect data from practitioners and answer the research questions.
Semi-structured interviews were selected as they are an effective approach to exploring participants' thoughts and experiences in depth \cite{dejonckheere2019semistructured}.
Regarding the interviewee selection, we aimed at recruiting participants that have different roles in the organization and have extensive experience in dealing with technical debt; these characteristics would allow us to explore the SATD management process and its tooling support from different perspectives (see \cref{sec:objective}).
Thus, the contact person at the company selected and invited the interviewees based on these characteristics from approximately 60 embedded software engineers.
These invitations were accepted by 12 practitioners.

Before the interviews, we extracted work artifacts from the selected project as aforementioned and identified SATD from them, as illustrated in \cref{fig:framework}.
Then we selected a sample of 15 SATD items, to show to the practitioner in order to help them gain a basic understanding of what automatically identified SATD looks like and prepare for the interviews.
The sample of 15 SATD items was selected from the set of identified SATD items in the previous step based on the proportion of different types of SATD, and consisted of 5 items from source code comments, 5 items from commit messages, and 5 items from issues.
For example, two of the presented SATD items were: \textit{``stupid code, why isn't this part of [function name]?''} and \textit{``adding sanity check on timing.''}
Besides, we provided practitioners with an introductory document on SATD (including a definition and examples of SATD as well as the high-level goal of this study).

\begin{table}[th]
\caption{Background information of interview participants.}
\label{tb:backgroud}
\begin{center}
\resizebox{\columnwidth}{!}{
\def\arraystretch{1.2}
\begin{tabular}{@{\extracolsep{4pt}}C{1.9cm}C{2.8cm}C{2.3cm}@{}}
\hline
Interviewee ID & Role in the Company & Years of Experience \\ \hline
\textit{I\textsubscript{1}} & Architect & 22 \\ 
\textit{I\textsubscript{2}} & Architect & 19 \\
\textit{I\textsubscript{3}} & Architect & 20 \\
\textit{I\textsubscript{4}} & Software developer & 22 \\
\textit{I\textsubscript{5}} & Software developer & 22 \\
\textit{I\textsubscript{6}} & Software developer & 20 \\
\textit{I\textsubscript{7}} & Software developer & 32 \\
\textit{I\textsubscript{8}} & Software developer & 17 \\
\textit{I\textsubscript{9}} & Team lead & 18 \\
\textit{I\textsubscript{10}} & Software developer & 34 \\
\textit{I\textsubscript{11}} & Team lead & 32 \\
\textit{I\textsubscript{12}} & Project manager & 24 \\
\hline
\end{tabular}
}
\end{center}
\vspace{-2mm}
\end{table}

Practitioners were then interviewed one by one by at least two of the authors via a web-based platform.
We asked practitioners to answer questions relating to their background, namely their role in the company and years of experience.
This background information is presented in \cref{tb:backgroud}.
After the background information collection step, we asked interviewees some introductory questions (e.g., What is your understanding of technical debt? Can you tell me some examples of technical debt?).
These ``warm-up'' questions encouraged interviewees to think about their own experiences with technical debt so that they can answer the rest of the questions based on those experiences.
During the interviews, we provided statistics on SATD (such as numbers and percentages of different types of SATD from different sources) in the selected project and the sample of identified SATD items.
Practitioners were asked to think about the SATD examples and statistics before answering interview questions.
Finally, the main part of the interview consisted of several questions aimed at  answering the Research Questions (as shown in \cref{tb:questions}); these were developed by following the interview guidelines of Seidman \cite{seidman2006interviewing}.
We asked the practitioners to talk about their ideas and opinions freely without restrictions.
During the interviews, we also asked follow-up questions to delve into their experiences and understanding.
Each interview took approximately 30 minutes.
After obtaining permission from interviewees, interviews were recorded to be transcribed for analysis.

\subsection{Data Analysis}

\subsubsection{Analysis of Work Artifacts}

To identify SATD from work artifacts, we first collect data from the selected projects, as discussed in Section III-C.
Subsequently, we followed the results of our previous work \cite{li2022automatic} to identify SATD from different sources using a deep learning approach.
This work is the only one focusing on accurately capturing SATD from different sources; specifically, the trained deep learning model achieved an f1-score (i.e., the harmonic mean of precision and recall) of 0.666, 0.644, 0.557 when identifying SATD from source code comments, commit messages, and issue tracking systems respectively.
Moreover, the machine learning model can identify four types of SATD, namely code/design debt, requirement debt, documentation debt, and test debt.
Examples of each type of SATD are presented in \cref{tb:satd_examples}.

\begin{table}[htb]
\caption{Examples of different types of SATD.}
\label{tb:satd_examples}
\begin{center}
\resizebox{\columnwidth}{!}{
\def\arraystretch{1.2}
\begin{tabular}{@{\extracolsep{4pt}}C{2.35cm}L{6cm}@{}}
\hline
Debt Type & Example \\
\hline
Code/Design debt & \textit{``Perl protocol handler could be more robust against unrecognised types''} - [from Thrift-issue] \\
 & \textit{``Need to add better handling for hz instance cleanup.''} - [from Camel-issue] \\
\hline
Test debt & \textit{``TODO: need more tests} - [from JMeter-code-comment] \\
 & \textit{``Tweaks tests to be a bit more robust''} - [from TrafficServer-commit] \\
\hline
Doc. debt & \textit{``FIXME: Document difference between warn and warning''} - [from JRuby-code-comment] \\
 & \textit{``we need to add it to the wiki page''} - [from Camel-issue] \\
\hline
Req. debt &  \textit{``TODO: add a dynamic context...} - [from Heron-code-comment] \\
 & \textit{``Union is not supported yet... I might be adding that capability quite soon.''} - [from Samza-pull] \\
\hline
\end{tabular}
}
\end{center}
\vspace{-2mm}
\end{table}

\subsubsection{Interviews}

To analyze the interviews, we first transcribed all interview recordings.
It is noted that one of the interviewees did not grant us permission to record the interview, so this interview was transcribed on the fly during the meeting.
Then, we followed an iterative qualitative data analysis process according to the \emph{Constant Comparative} method of \emph{Grounded Theory} \cite{strauss1990basics,stol2016grounded}.
Specifically, the analysis process is composed of three main steps.
The first step is \emph{open coding}, which breaks the transcript text down to discrete textual segments, which are subsequently coded (i.e. labeled).
When reading the interview transcripts, we continuously added new codes or changed current codes when necessary.
The scope of codes varies, as it could be a phrase, a sentence, or a paragraph.
Second, we applied \emph{selective coding}, by constantly comparing different codes and annotations, and then merging similar codes.
Third, we worked on the \emph{theoretical coding} to establish conceptual relations between codes.

To ensure the agreement on codes, the first and second authors independently performed the \emph{Constant Comparative} analysis process, discussed, and compared the generated codes to eliminate bias.
Any disagreements between the two authors were subsequently resolved.

We used a professional qualitative analysis tool (ATLAS.ti\footnote{\url{https://atlasti.com}}) to analyze the interview data.
The analysis results and interview protocol are available in the replication package\cref{l:data}.

\section{Results}
\label{sec:results}

\subsection{\textit{(RQ1.1) What Are the Types, Amounts, Resolution Time, and Sources of SATD Items in Industrial Settings?}}
\label{sec:td_characteristics}

\begin{table}[b]
\caption{Number of Different Types of SATD Items from Different Sources.}
\label{tb:num_of_td}
\begin{center}
\resizebox{\columnwidth}{!}{
\def\arraystretch{1.2}
\begin{tabular}{@{\extracolsep{4pt}}C{2.5cm}C{1.3cm}C{1cm}C{1.1cm}C{1cm}@{}}
\hline
\multirow{2}{*}{Debt Type} & \multicolumn{3}{c}{Source} & \multirow{2}{*}{Total} \\
\cline{2-4}
 & Comment & Issue & Commit &  \\
\hline
Code/Design debt & 3,139 & 9,318 & 2,236 & 14,693 \\
Req. debt & 602 & 702 & 119 & 1,423 \\
Doc. debt & 225 & 1,350 & 199 & 1,774 \\
Test debt & 63 & 540 & 93 & 696 \\
\hline
All SATD & 4,029 & 11,910 & 2,647 & 18,586 \\
\hline
\end{tabular}
}
\end{center}
\end{table}

\cref{tb:num_of_td} presents the number of different types of SATD from different sources.
We can observe that \textbf{most of the identified SATD is \emph{code/design debt} (79.1\%)}, followed by \emph{documentation debt} and \emph{requirement debt} (9.5\% and 7.7\% respectively).
The least amount of identified SATD is \emph{test debt} (3.7\%).

\begin{table}[htb]
\caption{Percentages of Different Types of SATD Items from Different Sources.}
\label{tb:percentage_of_td}
\begin{center}
\resizebox{\columnwidth}{!}{
\def\arraystretch{1.2}
\begin{tabular}{@{\extracolsep{4pt}}C{2.5cm}C{1.3cm}C{1cm}C{1.1cm}C{1cm}@{}}
\hline
\multirow{2}{*}{Debt Type} & \multicolumn{3}{c}{Source} & \multirow{2}{*}{Total} \\
\cline{2-4}
 & Comment & Issue & Commit & \\
\hline
Code/Design debt & \underline{2.0\%} & \textbf{12.8\%} & 10.7\% & 5.9\% \\
Req. debt & \underline{0.4\%} & \textbf{1.0\%} & 0.6\% & 0.6\% \\
Doc. debt & \underline{0.2\%} & \textbf{1.9\%} & 1.0\% & 0.7\% \\
Test debt & \underline{0.0\%} & \textbf{0.7\%} & 0.4\% & 0.3\% \\
\hline
All SATD & \underline{2.6\%} & \textbf{16.3\%} & 12.7\% & - \\
\hline
\end{tabular}
}
\end{center}
\end{table}

As mentioned in Section III-C, the issues come from all of the embedded software, while comments and commits are only from one (large) project.
Thus, we cannot compare the absolute numbers of SATD items directly between sources.
Thus, we look into the percentages of items across the different sources that contain SATD of different types (see \cref{tb:percentage_of_td}).
It is noted that, in this and subsequent tables, the highest values are highlighted in bold, while the lowest values are underlined.
Specifically, we calculate the percentages of different types of SATD by dividing the number of SATD items of a specific type from a specific source by the number of items from this source.
We observe that \textbf{the percentage of issues or commits being SATD issues or commits is significantly greater than source code comments} (16.3\% and 12.7\% versus 2.6\%). 
Finally, the percentage of issues being SATD is slightly greater than commits.

\begin{figure}[htb]
\centering
  \includegraphics[width=\linewidth, trim=10cm 10cm 10cm 5cm, clip=true]{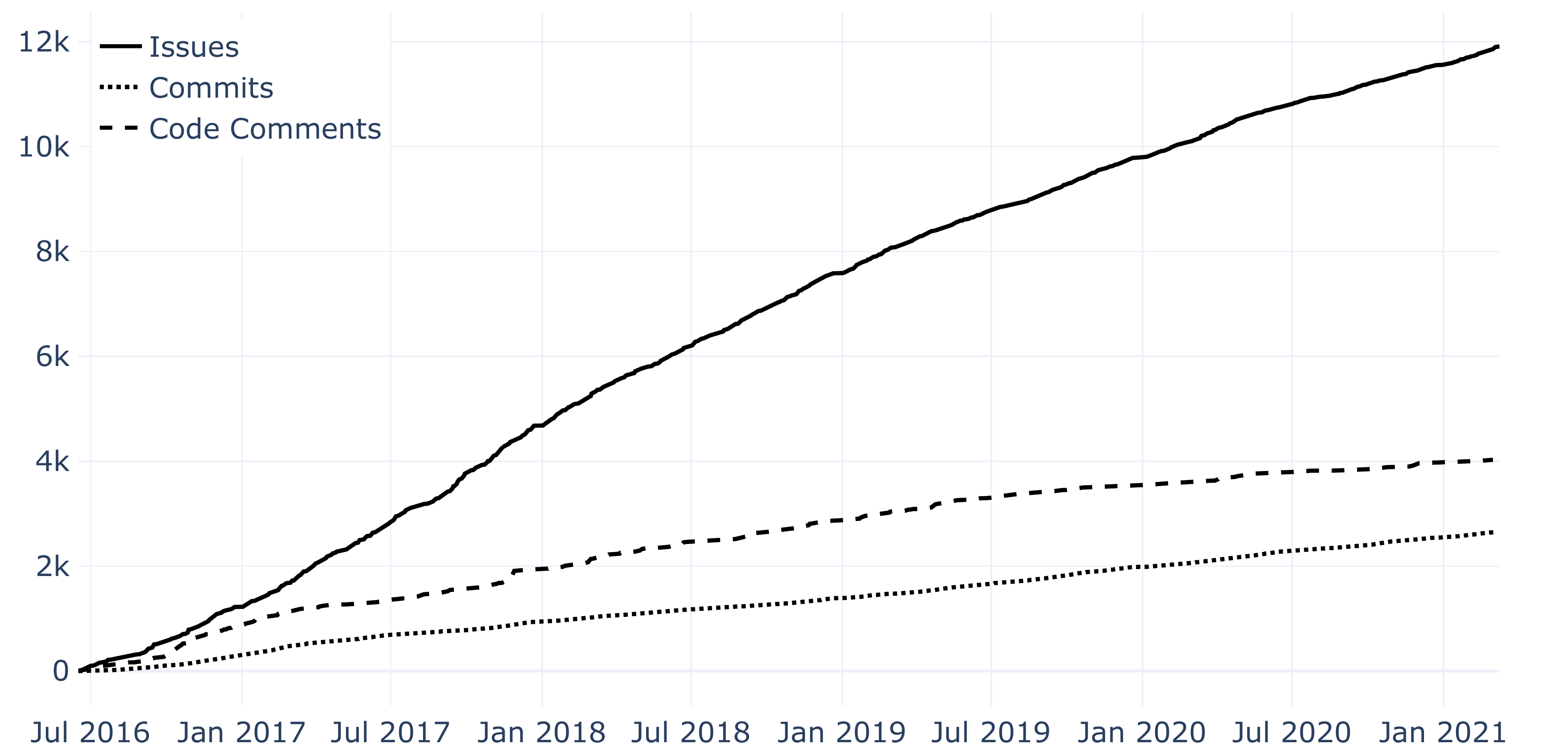}
  \caption{Cumulative number of SATD items in different sources over time.}
  \label{fig:num_td_over_time}
\end{figure}

\cref{fig:num_td_over_time} presents the number of cumulative technical debt admitted in different sources over time.
As can be seen, software developers keep documenting technical debt in different sources.
At the beginning of the studied period (before January 2017), the number of SATD in source code comments is comparable with the number of SATD in issues.
Afterward, the rate of admitting technical debt in issues increases compared to source code comments.

Because issue tracking systems provide additional information (e.g., issue type, issue status, issue closed time) that is related to SATD introduction and repayment, we further investigate SATD in issue tracking systems.
Specifically, we investigate the time required to resolve issues (with and without SATD), and the types of issues (e.g. backlog item or bug) with SATD. These are presented in the rest of this sub-section.

\cref{tb:time_and_pct_close} presents the average time to close different types of issues and the percentage of different types of issues that are closed.
As we can see, the average time to close different types of issues varies: \textbf{the average time to close non-SATD issues is shorter (47.2 days) compared to different types of SATD issues}; test debt issues take the longest average time to close (80.7 days).
Moreover, non-SATD issues achieve the highest closed rate (75.5\%), while requirement debt issues have the lowest closed rate (60.8\%).

\begin{table}[t]
\caption{Average Time to Close Issues and Percentage of Closed Issues.}
\label{tb:time_and_pct_close}
\begin{center}
\resizebox{\columnwidth}{!}{
\def\arraystretch{1.2}
\begin{tabular}{ccc}
\hline
Type & Avg. Time to Close (d) & Pct. of Closed (\%) \\
\hline
Code/Design debt & 62.5 & 71.3 \\
Req. debt & 70.2 & \underline{60.8} \\
Doc. debt & 60.4 & 72.0 \\
Test debt & \textbf{80.7} & 67.0 \\
Non-SATD & \underline{47.2} & \textbf{75.5} \\
\hline
\end{tabular}
}
\end{center}
\end{table}

\begin{table}[htb]
\caption{Comparison of Average Time to Close Issues Between Different Types of SATD and Non-SATD Issues.}
\label{tb:time_to_close}
\begin{center}
\resizebox{\columnwidth}{!}{
\def\arraystretch{1.2}
\begin{tabular}{@{\extracolsep{4pt}}R{2.6cm}L{2.05cm}C{0.95cm}C{1.6cm}@{}}
\hline
\multicolumn{2}{c}{Pairwise Comparison} & \textit{p}-value & Cliff's Delta \\
\hline
Code/Design debt & \multirow{4}{*}{\& Non-SATD} & \textbf{1.1e-20} & 0.12 (small) \\
Req. debt & & \textbf{1.3e-4} & 0.24 (small) \\
Doc. debt & & \textbf{9.9e-4} & 0.20 (small) \\
Test debt & & \textbf{3.3e-7} & 0.19 (small) \\
\hline
Code/Design debt & \multirow{4}{*}{\& Test debt} & \textbf{0.014} & 0.07 \\
Req. debt & & 0.361 & -0.06 \\
Doc. debt & & \textbf{0.020} & -0.01 \\
Non-SATD & & \textbf{3.3e-7} & -0.19 (small) \\
\hline
Code/Design debt & \multirow{4}{*}{\& Doc. debt} & 0.641 & 0.07 \\
Req. debt & & 0.160 & -0.06 \\
Test debt & & \textbf{0.020} & 0.01 \\
Non-SATD & & \textbf{9.9e-4} & -0.20 (small) \\
\hline
Code/Design debt & \multirow{4}{*}{\& Req. debt} & 0.247 & 0.12 (small) \\
Doc. debt & & 0.160 & 0.06 \\
Test debt & & 0.361 & 0.06 \\
Non-SATD & & \textbf{1.3e-4} & -0.24 (small) \\
\hline
Req. debt & \multirow{4}{*}{\makecell[l]{\& Code/Design\\debt}} & 0.247 & -0.12 (small) \\
Doc. debt & & 0.641 & -0.07 \\
Test debt & & \textbf{0.014} & -0.07 \\
Non-SATD & & \textbf{1.1e-20} & -0.12 (small) \\
\hline
\end{tabular}
}
\end{center}
\end{table}

Specifically, to evaluate the significance level and the effect size of closing time between different types of issues, we choose Mann-Whitney test \cite{mann1947test} and Cliff's delta \cite{vargha2000critique}.
The Mann-Whitney test is used to determine if two groups are significantly different from each other and is widely used in software engineering studies \cite{li2020identification,huang2018identifying}.
The results are demonstrated in \cref{tb:time_to_close}, while the \textit{p}-value is highlighted when it is less than $0.05$, which indicates the result has statistical significance.
We can notice that, \textbf{in contrast to previous research findings \cite{bellomo2016got}, there are significant differences between the closing time of non-SATD issues and different types of SATD issues} (\textit{p}-values are \num{1.1E-20}, \num{1.3e-4}, \num{9.9e-4}, and \num{3.3e-7} respectively).
Moreover, according to the Cliff's delta (i.e., 0.12, 0.24, 0.20, and 0.19 respectively), we can observe that the effect sizes\footnote{Effect sizes are marked as \emph{small} ($0.11 \le d < 0.28$), \emph{medium} ($0.28 \le d < 0.43$), and \emph{large} ($0.43 \le d$) based on suggested benchmarks \cite{vargha2000critique}} between them are all categorized as \emph{small} \cite{vargha2000critique}.
Furthermore, closing time differences between test debt issues and code/design debt issues or documentation debt issues are statistically significant (\textit{p}-values are 0.014 and 0.020).
However, their effect sizes are \emph{negligible} based on the Cliff's delta (i.e., 0.07 and 0.01).
As specified by the effect size in \cref{tb:time_to_close}, we find that the closing time differences between non-SATD issues and different types of SATD issues are greater than the closing time between different types of SATD issues.
Additionally, we compare the average time to close different types of issues in \cref{tb:avg_time_close_issue_type}.
As we can see, except for \emph{test issues}, all other types of issues align with the finding that SATD items take longer to solve.
The reason for \emph{test issues} not following this trend, might be due to the insufficient number of \emph{test issues} in comparison to other types of issues (247 vs. 885/4822/1333/4492).

\begin{table}[htb]
\caption{Average Time to Close Different Types of Issues.}
\label{tb:avg_time_close_issue_type}
\begin{center}
\resizebox{\columnwidth}{!}{
\def\arraystretch{1.2}
\begin{tabular}{@{\extracolsep{4pt}}C{2.35cm}C{1.0cm}C{1.0cm}C{0.8cm}C{0.8cm}C{0.8cm}@{}}
\hline
\multirow{2}{*}{Type} & \multicolumn{5}{c}{Issue Type} \\
\cline{2-6}
 & Feature & Backlog Item & Bug & Task & Test \\
\hline
SATD & 196.6 & 91.7 & 75.3 & 25.5 & \underline{897.6} \\ 
Non-SATD & \underline{186.6} & \underline{75.9} & \underline{68.0} & \underline{19.4} & 1110.0 \\
\hline
\end{tabular}
}
\end{center}
\end{table}

\begin{figure}[b]
  \centering
  \includegraphics[width=0.65\linewidth]{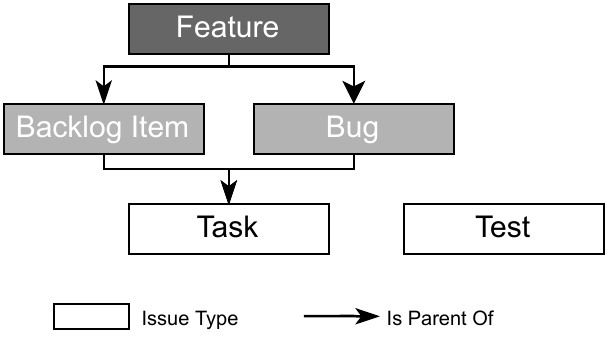}
  \caption{Hierarchy of issue types.}
  \label{fig:issue_type}
\end{figure}

Next, we study the occurrence of the types of SATD items (e.g. design or test debt) in the different types of issues (e.g. backlog or bug).
Issue tracking systems typically provide a function that helps developers categorize and track the progress of specific types of work \cite{issue_type}.
In the studied case company, the issue tracking system (Microsoft TFS) similarly supports specifying different types of issues.
The most common issue types used by the case company, are \emph{feature}, \emph{backlog item}, \emph{task}, \emph{bug}, and \emph{test}.
The hierarchy of issue types is illustrated in \cref{fig:issue_type}.
The studied organization uses these five issue types as defined by Microsoft \cite{issue_type2}: %
\emph{feature} is the highest-level type of work, it is associated with a specific product feature, and it is the parent of \emph{backlog item} and \emph{bug}.
\emph{Backlog item} is used to track development work, while \emph{bug} is for tracking code defects.
Moreover, \emph{task} is used to track fine-grained work, i.e. it is a child of both \emph{backlog item} and \emph{bug}. 
Additionally, test-related issue types are used independently of other types.
Because there are three test-related types, namely \emph{test case}, \emph{test plan}, and \emph{test suite}, we group them together under the category of \emph{test}.
\cref{tb:num_of_td_in_issues} presents the number of different types of SATD in accordance with these different issue types.
As we can see, most of SATD is identified as \emph{backlog item} and \emph{task} (4,822 and 4,492 respectively).
This indicates that \emph{backlog item} and \emph{task} are the two most popular issue types to admit technical debt.

\begin{table}[th]
\caption{Number of Different Types of SATD Items in Different Types of Issues.}
\label{tb:num_of_td_in_issues}
\begin{center}
\resizebox{\columnwidth}{!}{
\def\arraystretch{1.2}
\begin{tabular}{@{\extracolsep{4pt}}C{2.35cm}C{1.0cm}C{1.0cm}C{0.8cm}C{0.8cm}C{0.8cm}@{}}
\hline
\multirow{2}{*}{Debt Type} & \multicolumn{5}{c}{Issue Type} \\
\cline{2-6}
 & Feature & Backlog Item & Bug & Task & Test \\
\hline
Code/Design debt & 695 & \textbf{3,770} & 1,204 & 3,355 & \underline{220} \\
Req. debt & 58 & \textbf{350} & 37 & 211 & \underline{10} \\
Doc. debt & 85 & 512 & 46 & \textbf{701} & \underline{0} \\
Test debt & 47 & 190 & 46 & \textbf{250} & \underline{17} \\
\hline
All SATD & 885 & \textbf{4,822} & 1,333 & 4,492 & \underline{247} \\
\hline
\end{tabular}
}
\end{center}
\end{table}

\begin{figure}[htb]
\centering
  \includegraphics[width=\linewidth, trim=1.5cm 2cm 3.5cm 2cm, clip=true]{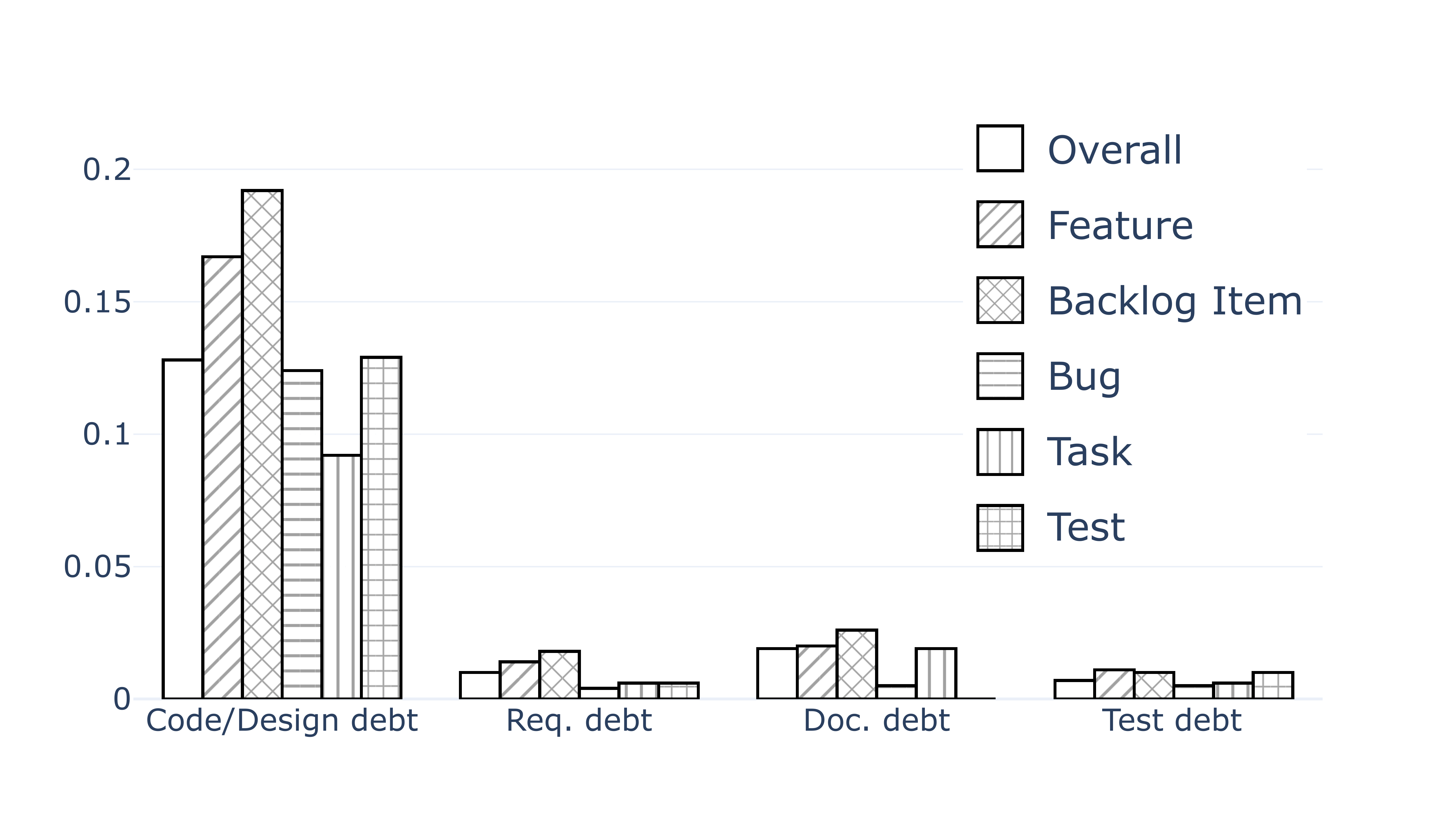}
  \caption{Percentage of Different Types of SATD Against Different Issue Types.}
  \label{fig:perc_td_against_issue_type}
\end{figure}

Because the total numbers of the different types of issues vary, it is unclear which issue type has the highest percentage of SATD issues. To address this, we show the percentage of different types of SATD in accordance with different issue types in \cref{tb:perc_of_td_in_issues,fig:perc_td_against_issue_type}.
We can notice that although \emph{backlog item} and \emph{task} have similar number of SATD issues (4,822 versus 4,492) in \cref{tb:num_of_td_in_issues}, \emph{backlog item} has a significantly higher percentage of SATD issues compared to \emph{task} (24.5\% versus 12.4\%).
This means that \textbf{\emph{backlog item} has the highest percentage and number of SATD issues among all issue types; in other words, it is the most used issue type for admitting technical debt}. This is in line with the definition of technical debt \cite{avgeriou2016managing}: while defects and poorly/partially implemented features are symptoms of technical debt, pure technical debt items concern issues that directly affect the maintenance and evolution of a software system. 

\begin{table}[ht]
\caption{Percentage of Different Types of SATD Items in Different Types of Issues.}
\label{tb:perc_of_td_in_issues}
\begin{center}
\resizebox{\columnwidth}{!}{
\def\arraystretch{1.2}
\begin{tabular}{@{\extracolsep{4pt}}C{2.35cm}C{1.0cm}C{1.0cm}C{0.8cm}C{0.8cm}C{0.8cm}@{}}
\hline
\multirow{2}{*}{Debt Type} & \multicolumn{5}{c}{Issue Type} \\
\cline{2-6}
 & Feature & Backlog Item & Bug & Task & Test \\
\hline
Code/Design debt & 16.7\% & \textbf{19.2\%} & 12.4\% & \underline{9.2\%} & 12.9\% \\
Req. debt & 1.4\% & \textbf{1.8\%} & \underline{0.4\%} & 0.6\% & 0.6\% \\
Doc. debt & 2.0\% & \textbf{2.6\%} & 0.5\% & 1.9\% & \underline{0.0\%} \\
Test debt & \textbf{1.1\%} & 1.0\% & \underline{0.5\%} & 0.6\% & 1.0\% \\
\hline
All & 21.2\% & \textbf{24.5\%} & 13.8\% & \underline{12.4\%} & 14.5\% \\
\hline
\end{tabular}
}
\end{center}
\end{table}

Additionally, as can be seen in \cref{fig:perc_td_against_issue_type}, \emph{feature} and \emph{backlog item} have a higher chance to contain code/design debt (16.7\% and 19.2\% compared to the average of 12.8\%), while \emph{task} only has 9.2\% of items being code/design debt (which is lower than average).
Moreover, the percentages of requirement debt for \emph{feature} and \emph{backlog item} are also higher than other types of issues (1.4\% and 1.8\%, respectively).
Finally, issues with the tags of \emph{bug} and \emph{test} are less likely to have documentation debt (0.5\% and 0\% compared to the average percentage of 1.9\%).

\subsection{\textit{(RQ1.2) How Is Automatically Identified SATD Regarded by Professional Software Engineers?}}
\label{sec:satd_attitude}

We report here the opinions of the interviewees on identified SATD and corresponding statistics produced by the automated SATD analysis.
First, we present the attitude towards SATD identified from different sources (two examples of identified SATD are shown in \cref{sec:approach}):

\begin{itemize}
    \item \textit{Attitude towards SATD identified from code comments.}
    We found that eight out of ten interviewees that commented on this, confirmed that SATD identified from code comments is indeed technical debt from their perspective: \textit{``yeah, those are the typical things [technical debt] that we enter in the code indeed.''}
    The other two interview participants also identified the vast majority of the discussed SATD items but were not very sure about one or two items: \textit{``the first one I would say difficult, it could also be a matter of taste; [...] the last one is the same as the first one, really depends on the situation.''}
    
    \item \textit{Attitude towards SATD identified from issues.}
    Six out of seven interviewees acknowledged SATD identified from issues as debt: \textit{``I expect them to be part of the backlog list, but I cannot explain to you one by one; I think they are technical debt.''} 
    Meanwhile, one interviewee found it difficult to judge whether it is SATD or not.%
    
    \item \textit{Attitude towards SATD identified from commits.}
    Seven out of eight participants confirmed that SATD identified from commits is technical debt from their point of view.
    Interestingly, four out of these seven participants pointed out that SATD in commits concerns documentation of paying back technical debt instead of incurring technical debt: \textit{``do you recognize technical debt in commits? yeah, but I think these are [documented] when somebody solves the technical debt in commit messages.''}
    There is one interviewee that did not acknowledge SATD in commits: \textit{``we always added text block in commits but not technical debt in commit messages.''}
\end{itemize}

Second, we discuss the participants' opinions about the importance of the identified SATD items.
Five out of nine interviewees mentioned that they need more information to determine the importance: \textit{``you have to know the implementation to have some insights on how severe such a thing is and how much work it will be to solve it; it [importance] is not immediately clear from the TODO itself.''}.
Besides, two out of nine participants believed that some of the items are not important, while the others need more information to evaluate: \textit{``the first one that I would say it's something that isn't really going to be resolved [...] the third one - it looks like it depends a bit on the on the functionality; is this really important or not, which is difficult to determine.''}
Meanwhile, the rest two of the nine interviewees pointed out that none of the identified SATD are important: \textit{``it does not have urgency to be solved.''}

Third, we describe the attitudes towards average time to close issues (see \cref{tb:time_and_pct_close,tb:time_to_close}).
Seven out of nine participants expected the same as the figure they were shown: \textit{``I think that is correct, which is about the balancing I told you between new functionality and technical debt.''}
They also mentioned that the reason behind this phenomenon is that they are under big pressure to implement new features or fix bugs instead of improving the quality of the code by solving SATD: \textit{``I know [this] project is in challenging phase; they are high pressured to reach the time-to-market, [so] we are also under pressure to have shortcuts and do not redesigns [unless we are] told necessary by the developers.''}
One interviewee agreed that technical debt items take a longer time to be resolved compared to non-debt items, but he also pointed out that he expected documentation debt to take the longest time to be solved among all types of debt: \textit{``I did not expect test debt takes that long; I would have expected the documentation debt to be there the biggest one.''}
Besides, one participant had different expectations than the results: \textit{``I think it's a matter of calculation; it's the other way around [compared to the expectation].''}

Fourth, we report the thoughts of participants towards all the presented statistics (see \cref{tb:num_of_td,tb:percentage_of_td,tb:num_of_td_in_issues,fig:num_td_over_time,tb:perc_of_td_in_issues,tb:time_and_pct_close,tb:time_to_close}) and identified SATD items (some examples are shown in \cref{sec:approach}):

\begin{itemize}
    \item \textit{Giving insights on how technical debt is managed.}
    Five interviewees indicated that statistics help them understand how technical debt is managed in their projects: \textit{``if you look at the statistics, I think that's the objective view of how things are managed.''}
    Furthermore, two of the participants considered it useful that the statistics provide information about the different types of SATD and the average time spent on SATD items and non-debt items.
    
    \item \textit{Showing what are the focus points.}
    One participant mentioned that statistics also show what the team emphasizes during SATD management: \textit{``do you think statistics is useful? yeah, [...] I would say [I know] what is focused on.''}
    
    \item \textit{Increasing the awareness of SATD.}
    Five interviews revealed that statistics and identified SATD help developers be conscious of technical debt in the projects: \textit{``it could always help to make us aware of technical debt.''}
    
    \item \textit{Providing insights on future improvement.}
    One interviewee stated that statistics could help developers become better and achieve higher productivity: \textit{``it gives some insight on how can you improve and be more efficient in your work.''}
\end{itemize}

\subsection{\textit{(RQ1.3) What Are the Relations Between SATD in Different Sources?}}
\label{sec:relations}

The relations between SATD items in different sources, as derived from the data, are summarized and presented in \cref{fig:relations}.
It is evident that technical debt admitted in code comments and issues is referenced in the other two sources, while technical debt admitted in commits is not mentioned in other sources.
Because each issue has a unique issue ID, developers can refer to that ID when referencing a SATD issue: \textit{``in most cases, we try to add the issue ID within the comments.''}
We call this reference a \emph{specific reference}.
On the other hand, since there is no unique identifier for each source code comment, it is impossible to reference specific comments.
Thus, such references are usually \emph{approximate references}.
We describe the relations in detail as they are numbered in \cref{fig:relations}:

\begin{figure}[htb]
\centering
  \includegraphics[width=\linewidth]{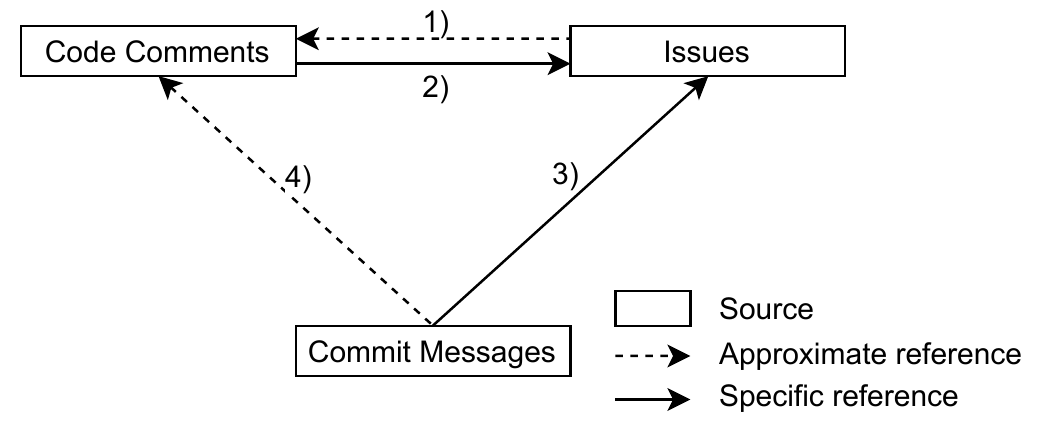}
  \caption{Relations between SATD items in different sources.}
  \label{fig:relations}
\end{figure}

\begin{enumerate}
    \item \textit{Technical debt admitted in code comments is referenced in issues.}
    Two interviewees mentioned that it is not common to reference SATD code comments within issues: \textit{``in the issues, nine out of ten times, [developers] never write down which actually line or file [is] related.''}
    However, one of them also pointed out that developers sometimes note down in issues the approximate location of technical debt which has been documented in code comments: \textit{``[developers] only specify a certain piece of code where the problem resides.''}
    
    \item \textit{Technical debt admitted in issues is referenced in code comments.}
    The links from issues to code comments are mentioned by four interview participants.
    They tend to add issue IDs (unique identifiers) in the code comments to establish clear links: \textit{``sometimes you add a link to the issue in the code.''}
    
    \item \textit{Technical debt admitted in issues is referenced in commit messages.}
    Three interviewees mentioned that SATD in issues is also referenced in commits: \textit{``in the commit, we are able to tag the issue item, and then the link between commits and issues is made automatically.''}
    This gives developers a better understanding of the changes in the commits: \textit{``I do not know if it was a single line commit message, which is vague, short, and without explanation [...] we need to have more information in the commit or a link to the issues.''}
    
    \item \textit{Technical debt admitted in code comments is referenced in commit messages.}
    One interviewee indicated that the repayment of SATD in code comments might be documented in commit messages: \textit{``there could be a link if you have the previous one in comments, when somebody solved it, probably in the commit message you might record the resolve of it.''}
\end{enumerate}

\subsection{\textit{(RQ2.1) When Is Technical Debt (Not) Admitted in Source Code Comments, Issue Tracking Systems, and Commit Messages?}}
\label{sec:timepoint_doc_td}

During the interviews, we established that software engineers tend to admit technical debt in different sources for different reasons.
For each source (i.e., source code comments, issue tracking systems, and commit messages), we report several cases why technical debt is being admitted.
We start first with the \emph{source code comments}:

\begin{itemize}
    \item \textit{Scale of technical debt is small.}
    Four interviewees mentioned that developers tend to document small technical debt items in source code comments: \textit{``if it [technical debt] is too small, just admit it in the code comments.''}
    Regarding what \emph{small} technical debt actually means, as  an interviewee stated: \textit{``if you look at things in source code, they are typically smaller; those things are just magic number or making this as parameter...''}
    
    \item \textit{Solving technical debt brings little or no benefit.}
    When solving technical debt yields small or negligible benefit, developers tend to document it in comments: \textit{``if we will not gain the advantage over anyway, then probably something will be noted in the software [...] a comment will be added.''}
    
    \item \textit{Deciding not to fix the technical debt.}
    Two participants pointed out that if developers reach an agreement on not fixing the technical debt, they usually just document it in code comments: \textit{``if we already decide not to fix this technical debt, then probably it will remain as comments.''}
    
    \item \textit{Helping other developers to become familiar with technical debt related to code and its rationale.}
    Five interviewees mentioned that it is important to document technical debt and its rationale to help other developers become aware of problematic code and the reasons behind it: \textit{``the comments in software to make sure that when people are facing troubles and having a look at the software again that they know about the facts that have been made to some different choices and which could result in a problem.''}
    
    \item \textit{It concerns requirement debt.}
    Three interviews revealed that if the technical debt is of the type requirement debt, such as partially implemented requirements, developers prefer to document it in source code comments: \textit{``because we have not finished yet, it [code comment] is typically written down while developing the feature.''}
\end{itemize}

Second, for \emph{issue tracking systems}, technical debt is documented in the following cases:

\begin{itemize}
    \item \textit{Scale of technical debt is big.}
    Six interview participants indicated that developers always document large-scale technical debt in issue trackers: \textit{``If you look at issue tracker... you have to fix this entire piece of code, that's a bigger span, while in code it's basically for the next line.''}
    
    \item \textit{Technical debt is part of the future plan.}
    Five interviewees pointed out that developers always document technical debt in issues when they actually plan to fix them in the future: \textit{``I think that we create issues for them [technical debt] to make sure that they will become part of the future plans.''}
    
    \item \textit{Features only supported by the issue tracker.}
    Issue tracking systems provide features that are not provided by code comments or commits, such as uploading attachments and assigning severity levels for issues.
    An interviewee mentioned that he always summarizes technical debt and designs in a word document on a daily basis, then uploads it as an attachment when creating a new issue.
    
    \item \textit{Track technical debt repayment in the engineering phase.}
    Developers in this case study refer to software development in later iterations with \textit{engineering phase}, which is different from the early phase of development.
    When developers want to track what changes will be made to solve technical debt in the engineering phase, they create issues: \textit{``I think in engineering phase [when I] have to clean something up, I will definitely make issues, so you can always see what has been done; in the early phase, it's just about what works are expected.''}
    
    \item \textit{Duplicate existing technical debt admitted in code comments.}
    Three software engineers believed that existing technical debt in comments should also be admitted in issues to facilitate their tracking: \textit{``if there are still TODOs in the code, there should also be an issue that something still needs to be done.''}
\end{itemize}

Third, developers document technical debt in \emph{commit messages}, in the following cases:

\begin{itemize}
    \item \textit{Commits introducing Technical Debt.}
    One interviewee pointed out that, if commits include workarounds that are typical of technical debt, they usually document those problems, as well as the related issue keys in the commit messages: \textit{``we have certain commits which indicate that we have to make shortcuts or we have implemented a temporary situation.''}
    
    \item \textit{Commits related to technical debt repayment.}
    Three interviews disclosed that if commits are about technical debt repayment, they always document it in commit messages: \textit{``these are when somebody solved technical debt in these commit messages.''}
\end{itemize}

Finally, we also summarize the cases when technical debt is ignored or not documented in artifacts:

\begin{itemize}
    \item \textit{Developers are under pressure and forget to document technical debt.}
    Three participants pointed out that if the pressure is very high, developers usually focus on other work and postpone technical debt documentation; in most cases, it is eventually forgotten: \textit{``if the pressure is really high, [...] you will do that [technical debt] tomorrow, and tomorrow has another thing that got forgotten.''}
    
    \item \textit{Certain types of technical debt are ignored.}
    According to four interviews, some developers do not consider certain types of technical debt to be important and choose not to document them. 
    Specifically, interviewees believed that developers pay less attention to documenting test debt: \textit{``I think we don't have that many technical debt items for missing test cases; I think you more or less know about them, but no real documentation about test cases and actual implementation.''}
    
    \item \textit{Scale of technical debt is small.}
    When the scale of technical debt is small, developers may decide not to document it at all because of its low impact: \textit{``what are the reasons not documenting technical debt? [it depends on] how big is the technical debt, if just a small thing, it's probably not.''}
    
    \item \textit{Technical debt in legacy code.}
    Two interviewees reported that developers are aware of the limitations of technical debt in old parts of the system and choose not to document it, because they know it will not be fixed anyway: \textit{``[if] the architecture is already fifteen years old [...], you know what the limitations are, you can still write technical debt to make it better, but you know it will not be fixed anyway.''}
    
    \item \textit{Short life of technical debt.}
    We noticed that when developers think the technical debt will be solved in the near future, they might choose not to document it (mentioned by three interviewees): \textit{``I know that for the old release, we make a quick workaround, but we don't mark [it] as technical debt because we make the actual good solution in our mainline immediately.''}
    
    \item \textit{Direction is unclear in early phases.}
    Because of uncertainties in the early phases of projects, software engineers may choose not to document it: \textit{``At the beginning of the project, it can go anywhere, so if you put a lot of effort in explaining why something is done, it takes lots of time.''}
    
    \item \textit{The responsibility of other developers.}
    In some cases, developers are in charge of certain parts of software development or documentation update.
    When other developers encounter technical debt, they prefer to let the responsible person document it: \textit{``if it's someone else's documentation, I might mention it to someone. I usually do not create an issue for that.''}
    
    \item \textit{Treating technical debt as common knowledge.}
    One participant mentioned that technical debt is not documented when it is known by everyone in the team: \textit{``[technical debt] is not documented [...] [when we] have accepted [it], and treat [it] as a common knowledge of the team; the team members know that issue is there, or inconvenience is there.''}
\end{itemize}

\subsection{\textit{(RQ2.2) What Are the Pros and Cons of Admitting Technical Debt in Different Sources?}}
\label{sec:pro_con}

The pros and cons of documenting technical debt in \emph{source code comments} are summarized below:

\begin{itemize}
    \item \faThumbsOUp~\textit{Pointing to problems in the code.}
    Five interview participants pointed out that technical debt documented in code comments could help developers understand the existing technical debt in code and potentially solve it: \textit{``make sure that people are looking at the software [reading source code], they will be familiar with the fact that there is technical debt in code.''}
    
    \item \faThumbsOUp~\textit{Long lifetime of code.}
    One interview participant indicated that code is a very stable artifact compared to others.
    In contrast to other artifacts, comments in source code will not disappear in the future: \textit{``I have seen tools coming and gone; five years back we [switched the issue tracker] [...], [but] I have code older than five years, maybe ten years old, so I don't know the change request anymore from seven years back and the rationale; the only thing I have is just the source code.''}
    
    \item \faThumbsOUp/\faThumbsODown~\textit{Limited visibility.}
    On the one hand, documenting technical debt in code comments causes less disturbance to other developers:
    \textit{``too many detailed tasks [in issues] does not help which could bother teammates [...] just admit them in the code comments, [...] because you intend to solve it soon anyway.''}
    On the other hand, it could restrict the visibility of technical debt, resulting in paying less attention to it: \textit{``they are not visible anymore, only if you run into that.''} 
    
    \item \faThumbsODown~\textit{Resolving it depends on the initiative of developers.}
    Three interviewees reported that it highly depends on software developers to solve or leave technical debt admitted in code comments: \textit{``you need be lucky that someone will be working on this to get a solution.''} Thus, documenting technical debt in the source code can act both as an advantage (if it gets resolved) and a disadvantage (if it is ignored).
\end{itemize}

Subsequently, we list the main pros and cons of admitting technical debt in \emph{issues}:

\begin{itemize}
    \item \faThumbsOUp/\faThumbsODown~\textit{Visible to the whole team.}
    Six interviewees mentioned that technical debt admitted in issues has the advantage of being visible to everyone in the team, helping developers to keep track of it: \textit{``issue tracker is used for recording important technical debt which is shared in the team.''}
    
    \item \faThumbsOUp~\textit{Issue trackers provide features not supported by other artifacts.}
    Two interviewees revealed that issue tracking systems provide several features that support technical debt management.
    Specifically, issue trackers can give developers an overview of all documented technical debt: \textit{``it's a good thing that technical debt is mostly recorded in the issue tracker because this gives an overview.''}
    It also supports uploading technical debt information as attachments, assigning the issue type, and assigning the issue severity.
    Finally, it supports keeping track of what has been done about the technical debt: \textit{``I will make an issue, so you can always see what has been done.''}
    
    \item \faThumbsOUp~\textit{Support planning to resolve technical debt.}
    Six interviewees reported that technical debt documented in issues will be a part of the future plan and be resolved eventually.
    This is because, in addition to developers, team managers also participate in the management of SATD in issues: \textit{``[as the team lead] once they are in issues, they are in my list of choosing priorities, that I can deal with it.''}
\end{itemize}

Finally, the pros and cons of recording technical debt in \emph{commits} are presented below:

\begin{itemize}
    \item \faThumbsOUp~\textit{Providing explanation for TD changes.}
    Two interviewees mentioned that commits are important to explain what TD changes are made to the repository, such as introducing TD, modifying TD, and repaying TD: \textit{``commit messages should include what has changed and what has been done, also for changes to technical debt.''}
    
    \item \faThumbsOUp/\faThumbsODown~\textit{Limited visibility.}
    Similar to code comments, technical debt admitted in commits has limited visibility.
    From the viewpoint of the team lead, it is not visible to him: \textit{``if they are in comments or commits, they're not on my desk.''}
    
    \item \faThumbsODown~\textit{Resolving it depends on the initiative of developers.}
    There is no guarantee that technical debt admitted in commits will be resolved.
    It depends on the developers to solve it or leave it.
    The team lead only manages technical debt documented in issues: \textit{``I don't manage technical debt in code comments and commits at all, that's really depending on the engineer's responsibility.''}
\end{itemize}

\subsection{\textit{(RQ2.3) What Are the Triggers to Pay Back or Not Pay Back SATD?}}
\label{sec:trigger_repay}

According to the interview responses, software developers tend to repay SATD in the following cases:

\begin{itemize}
    \item \textit{SATD is involved in upcoming changes.}
    Based on six interviews, developers always choose to repay SATD when changes are going to take place in the same part of the system.
    This is because technical debt could make the changes more difficult: \textit{``for instance the parameterize thingy, if I was doing a change which actually needs that or in the same area, I would take that along because that would really help me if I solve it.''}
    
    \item \textit{SATD is related to bugs.}
    In another case, three interviewees reported that when they find SATD connected to bugs, they will solve the technical debt: \textit{``we really have to start solving [the technical debt] that keeps bugs popping up with that same piece of code,  [for example if] you have these bugs popping up [while] you see test debt, [it happens because you] don't have test cases in that area.''}
    
    \item \textit{SATD is experienced by stakeholders.}
    One interviewee indicated that SATD observed by stakeholders is more important: \textit{``I will focus first on technical debt that is experienced by stakeholders.''}
    
    \item \textit{SATD hinders other tasks.}
    Two participants pointed out that they need to repay SATD when it prevents them from keeping making progress: \textit{``if they are hindered by [technical debt], then it's important to focus on the bad choices.''}
    
    \item \textit{Small SATD that can be solved easily.}
    Based on three interviews, we found that when developers encounter SATD and think the debt can be paid back easily, they prefer to solve it straight away: \textit{``if there is a small [technical debt], there is time left in your sprint then you could pick up such a small item.''}
    
    \item \textit{The same SATD keeps annoying developers.}
    Two responses indicated that when developers encounter a technical debt item, which is repeatedly of concern to them, they will take some time to solve it: \textit{``if you hit the same technical debt item and it annoys you enough, then it will be solved.''}
    
    \item \textit{Certain types of SATD are valued more than others.}
    Some developers believe that certain types of SATD are more important than others, and they choose to repay them with higher priority.
    More specifically, two interviewees stated that they prioritize test debt: \textit{``I would prioritize test debt; that's critical on the code quality.''}
    On the other hand, two participants mentioned they always give design debt higher priority: \textit{``you also see preferences more to the design debt to documentation debt.''}
    
    \item \textit{Too much SATD in the area.}
    Five participants pointed out that when too much SATD is accumulated in a specific part of the system, it gets to be paid earlier rather than later: \textit{``if there is a lot of technical debt in those modules, you might want to pick up earlier, cause if there is some TD there, maybe something wrong in the design...''}
    
    \item \textit{Location of SATD is special.}
    One interview participant mentioned that SATD in different parts of the project is treated differently.
    They always give high priority to SATD in certain modules: \textit{``it is up to the part of the project if I make a shortcut that should not be in; I am aware of it and will resolve it.''}
    
    \item \textit{Potential risk of SATD is high.}
    According to three interviews, when the potential risk of SATD is very high, SATD should be worked on: \textit{``I think you should work on the most important things, the highest risk things.''} 
    
    \item \textit{Have sufficient time.}
    Three interviewees indicated that when they have sufficient time, they will take some time to solve SATD: \textit{``when do I decide to solve technical debt apart? when I have time in the program.''}
    
    \item \textit{Software craftsmanship.}
    Two interview participants reported that some developers have the attitude of striving to deliver software of high quality: \textit{``[if] I have craftsmanship, I deliver software as the input and believe the software needs to be correct and needs to be maintainable.''}
\end{itemize}

Meanwhile, in the following cases, SATD is ignored and left unresolved:

\begin{itemize}
    \item \textit{Repaying SATD brings small benefit.}
    Three interviewees mentioned that if the software works, repaying SATD yields only a small benefit. Thus, developers tend to not pay back the debt: \textit{``if it already works, why make it better? someone pays for it.''}
    
    \item \textit{Repaying SATD takes too much effort.}
    Four participants had concerns about the considerable effort required to pay back SATD: \textit{``we don't want to invest in it, due to [...] too much effort.''}
    
    \item \textit{Potential risks of paying back SATD.}
    Four interviewees expressed the concern that it could be risky to repay SATD, as it might break existing functionalities: \textit{``there is some regression risk involved; it should be simple but sometimes takes a long time to finish.''}
    
    \item \textit{Certain types of SATD are ignored during repayment.}
    As one of the participants mentioned: \textit{``the one writing [documentation] typically isn't the user of it''}; some developers simply give documentation debt low priority: \textit{``I don't like writing documentation, so I really try to postpone writing the document.''}
    Besides, two participants stated that some developers do not see test debt as important and choose not to repay it: \textit{``[some] engineers don't see writing tests really helps them because you have implemented something; it runs on a machine and it works.''}
    Moreover, another interviewee pointed out that architecture debt is sometimes ignored in the maintenance phase: \textit{``architecture debt is addressed differently than design and test debt because it more prevents production; architecture debt also depends on the development phase; if it goes to maintenance phase from earlier phases of building a new product, we often keep the debt as long as it doesn't break functionality.''}
    
    \item \textit{Learning effect for the SATD creator.}
    Another interviewee believed that the debt creator should solve it, to be able to learn from it: \textit{``it is important to close the feedback loop; if others resolve it [technical debt], the people who created it will never learn from it.''}
    
    \item \textit{Inactive code.}
    Two interviewees reported that when the code is inactive and there are no changes planned for it, related technical debt does not have priority to be paid back: \textit{``If there are no changes or features, or plans for this, maybe [it] will not be used anymore, then that's not so important.''}
    
    \item \textit{Careless developers.}
    Lastly, one interview revealed that irresponsible developers also lead to SATD unsolved: \textit{``some people say we should solve it, and then they don't stick to it.''}
\end{itemize}

\subsection{\textit{(RQ2.4) What Practices Are Used to Support SATD Management in Industrial Settings?}}
\label{sec:satd_management}

In the following, we summarize practices used to assist in SATD management.
First, we describe practices that help prioritize SATD using different criteria:

\begin{itemize}
    \item \textit{Custom list.}
    Four interviewees indicated that they maintain a list of SATD with an order of priority.
    Specifically, they usually put the high-priority SATD on the top of the list for quicker repayment: \textit{``[we] try to prioritize the list, and the most important items are on the top that needs to be solved first.''}
    
    \item \textit{Severity level of tickets.}
    Issue tracking systems always support setting the priority for each issue ticket (e.g., block, minor, and trivial) \cite{issue_priority}.
    One interview participant mentioned they also use the issue tracker's built-in function to set the priority of each issue: \textit{``most items are already categorized with a severity.''}
    
    \item \textit{Type of tickets.}
    Five interviewees mentioned that issue types have an impact on the priorities of issue tickets.
    They choose different issue types when creating issues with different priorities. For example, the interviewees considered that \emph{bugs} have higher priority than \emph{backlog items}: \textit{``I would say I had a \underline{task} if it is for short term if you intend to solve it within this sprint, if [...] you create a \underline{backlog} item, [it could just] disappear, so it would be better to write the \underline{bug} then at least you have a process to handle these.''}
    
    \item \textit{Referencing issue keys.}
    One participant indicated that when adding a reference to an issue ticket, the SATD in code comments will get higher priority: \textit{``if that comment references an issue, it will automatically get more priority.''}
\end{itemize}

Second, we report two common practices to efficiently pay back SATD:

\begin{itemize}
    \item \textit{Grouping related technical debt items.}
    Two participants indicated they usually group related technical debt items and investigate them together: \textit{``we group them [technical debt] together; that's we say, those four or five items are in the same area, let's now take a look at them together, to be more effective.''}
    
    \item \textit{Grouping technical debt and development tasks.}
    Two interviews revealed that developers also group technical debt and development tasks (e.g., fixing bugs, creating new features, and adding tests).
    Then they solve them jointly: \textit{``when we take technical debt we also resolve other things, which is more efficient.''}
\end{itemize}

\subsection{\textit{(RQ3.1) What Challenges Do Software Practitioners Face When Managing SATD?}}
\label{sec:challenges}

In the following, we summarize the challenges for SATD management:

\begin{itemize}
    \item \textit{Convincing developers not to introduce SATD, when not necessary.}
    Three interviewees indicated that some technical debt can be paid back easily, so it should not be incurred in the first place: \textit{``many of these [comments] seem to be fixed in five minutes, I think they shouldn't write these comments; they do not look like effort-intensive.''}.
    
    \item \textit{Prioritizing SATD.}
    Five interviewees pointed out that it is hard to determine priorities of SATD and other works: \textit{``the biggest challenge is setting the priorities [...] the challenge is always what's the best to do, a piece of functionality or technical debt?''}
    
    \item \textit{Getting resources to pay back SATD.}
    Based on two interviews, we found that getting resources for debt repayment remains challenging: \textit{``to get technical debt on the agenda is a difficult task [...] there's always an argument to not work [on them].''}
    
    \item \textit{Dealing with undocumented technical debt.}
    Three interview participants mentioned the difficulty of dealing with undocumented debt: \textit{``we struggle with the ones [technical debt] we are not aware of or somebody identified without clearly communicated as being technical debt.''}
    One interviewee specified that it is especially challenging when dealing with technical debt in old parts of the system without documentation: \textit{``what challenges did you face when dealing with technical debt? dealing with legacy code in general [...] re-engineering the code or design sometimes is difficult [...] it could be a lot of helpful if there is some code documentation.''} 
    
    \item \textit{No guideline for SATD documentation.}
    Four interviewees pointed out the problem of not having concrete guidelines for SATD documentation: \textit{``we don't have any agreements on when you have technical debt and want to add some comments within the software, then please use certain tags.''}
    
    \item \textit{No guideline for SATD repayment.}
    Besides, two interviewees reported that there is no guideline for repaying SATD: \textit{``there is no complete guideline; if you have to solve this, then you should do this, this, and this.''}
    
    \item \textit{Dealing with consequences of SATD.}
    Two interviewees found that it is challenging to deal with an increasing list of SATD items, as it causes SATD items to be ignored or not to document new technical debt: \textit{``sometimes the technical debt items get out of sight because the list is becoming too long and you forgot about it; I am not sure how to deal with that growing list of technical debt items.''}
    Meanwhile, another participant stated the harmfulness of accumulating technical debt without proper management: \textit{``in another project, it was horrible; the big redesigns block the whole development, which also affects the trust of the software.''}
\end{itemize}

\subsection{\textit{(RQ3.2) What Features Should Tools Have to Effectively Manage SATD?}}
\label{sec:tools}

In the following, we report the tool features that developers thought were useful for SATD management.
We categorize features into four groups: SATD identification, SATD traceability, SATD prioritization, and SATD repayment.
SATD identification-related features are reported as follows:

\begin{itemize}
    \item \textit{Automated SATD identification.}
    Seven interviewees mentioned that it would be useful to be able to automatically identify SATD from different sources: \textit{``I think [the tool should support] the identification of technical debt, scanning code or issues.''}
    The interviewees suggested the following ways to present the identified SATD:
    
    \begin{itemize}
        \item \textit{Show the list of identified SATD.}
        Two interviews indicated that identified SATD items should be listed: \textit{``if the tool could make some kind of printouts of technical debt items in your source code, then I can imagine that I will sit together with some engineers, walk through the list, find the most important, and solve them.''}
    
        \item \textit{Show the quantity of SATD in the system.}
        One participant suggested showing the number of SATD items in the dashboard to increase the visibility of SATD in code: \textit{``this dashboard will show you how much TODO in our code, so that is visible for the whole team.''} 
        
        \item \textit{Show the evolution of SATD quantity over time.}
        Another participant mentioned that it would be useful to have a function showing the number of SATD over time to know when they introduce more SATD or less SATD: \textit{``[the tool should show] total amount technical debt in the system evolving during the development of the system, [so we can know] do we have more technical debt in the early phase.''}
        
        \item \textit{Show the quantity of SATD in different modules.}
        As mentioned in \cref{sec:trigger_repay}, if too much SATD is accumulated in a part of the system or in some specific modules, developers would give the debt higher priority.
        Thus, they would like the tool to show the quantity of SATD in different modules: \textit{``what would you like to see? [I want to have] some insights into which module has a lot of technical debt.''}
    \end{itemize}
    
    \item \textit{Automated differentiation between fixed SATD and unfixed SATD.}
    As stated in \cref{sec:timepoint_doc_td}, the identified SATD may be either repaid or not. 
    There needs to be a distinction between them; as one participant  stated: \textit{``I am curious only about the open ones [unsolved debt].''} 
    The participants mentioned the following means to visualize the distinction between solved and unsolved SATD:
    
    \begin{itemize}
        \item \textit{Show the period between SATD introduction and repayment.}
        One interviewee mentioned that he wanted to know the repayment time of SATD: \textit{``[the tool should show] how much time is in between when we decided to introduce technical debt and when will things be solved.''}
        
        \item \textit{Show how long unsolved SATD survives.}
        Another interviewee indicated that the tool should show the survival time of SATD: \textit{``[the tool should show] how long technical debt is there.''} 
        
        \item \textit{Show the timeline of fixed and not fixed SATD items.}
        Another interviewee was interested in the point in time when they decide to either pay the TD back or not: \textit{``[the tool should show] what is the moment we solve most technical debt? when do we decide to leave technical debt in the system and stop working on them?''}
    \end{itemize}
\end{itemize}

Next, SATD traceability-related features are presented:

\begin{itemize}
    \item \textit{Automated tracing between SATD in different sources.}
    In \cref{sec:relations}, we observed some relations between SATD in different sources.
    But, some of these relations (e.g., technical debt admitted in code comments referenced in issues) are rarely documented.
    Thus, two interviewees think it would be very helpful if the tool could build traces automatically between SATD in different sources: \textit{``linking back and forth would really help in getting an overview about technical debt things.''}
    
    \item \textit{Automated tracing between SATD and code.}
    Two participants mentioned that it would be useful to know the location of SATD in the code: \textit{``I would like to know which area of code the technical debt is located at.''}
    
    \item \textit{Automated tracing between SATD and related development tasks.}
    As described in \cref{sec:trigger_repay}, when SATD is involved in upcoming changes, it is usually prioritized.
    Thus, developers are interested in which to-do items (e.g., fixing bugs, creating new features, or adding tests) are related to the SATD: \textit{``[the tool should find] related work to it.''}
\end{itemize}

Subsequently, we report the suggested features related to SATD prioritization:

\begin{itemize}
    \item \textit{Automated SATD prioritization.}
    Two interviewees wanted to automatically prioritize SATD: \textit{``I'm looking to which part of technical debt could be left and which part of technical debt really needs to pay attention to.''}
    
    \item \textit{Automated identification of SATD risk.}
    Three participants mentioned that SATD risk identification should be supported by the tool: \textit{``I am looking at [...] what is the pain? do I need to solve it? what are the consequences [of] not solving it?''}
    
    \item \textit{Automated estimation of benefits to solve SATD.}
    Two interviewees indicated that estimating the benefits of solving SATD (e.g. how much technical debt interest will be saved) could be one of the tool's functions: \textit{``need to know what the benefit of it [solving SATD], what is gained by it.''}
    
    \item \textit{Automated estimation of cost to solve SATD.}
    Based on four interviews, developers mentioned that automated estimation of SATD repayment cost (also referred to as the principal of technical debt) is useful: \textit{``the other thing is how much effort does it take to get rid of this technical debt.''}
\end{itemize}

Finally, there is one feature related to SATD repayment:

\begin{itemize}
    \item \textit{Automated SATD solution suggestion.}
    Two interview participants asked for the tool to provide some potential solutions (e.g. refactorings) for paying back SATD: \textit{``the other tool could [provide] [...] possible routes of solution.''}
\end{itemize}

\section{Discussion}
\label{sec:discussion}

According to the study design (see \cref{sec:approach}), we formulated three main research questions to investigate the nature of SATD, the SATD management activities, and SATD management improvement.
Thus, we organize the discussion into three parts: the discussion about the nature of SATD, the discussion about SATD management activities, and the discussion about SATD management improvement.

\subsection{Nature of SATD}

As mentioned in \cref{sec:approach}, there are significant differences between open-source and industrial projects.
It is important to know these differences in order to understand how to better manage SATD in the two cases: what works for each case, what works for both, and what can be reused from one to the other.
Thus, we compare the characteristics of SATD in industrial and open-source projects.

\begin{table}[htb]
\caption{Comparison Between Percentages of Different Types of SATD Items in Industrial Projects (IP) and Open-Source Projects (OSP).}
\label{tb:satd_perc_comparison}
\begin{center}
\resizebox{\columnwidth}{!}{
\def\arraystretch{1.2}
\begin{tabular}{@{\extracolsep{4pt}}ccccccc@{}}
\hline
\multirow{3}{*}{Debt Type} & \multicolumn{6}{c}{Source} \\
\cline{2-7}
 & \multicolumn{2}{c}{Comment} & \multicolumn{2}{c}{Issue} & \multicolumn{2}{c}{Commit} \\
\cline{2-3}
\cline{4-5}
\cline{6-7}
 & OSP & IP & OSP & IP & OSP & IP\\
\hline
Code/Design debt & 80.6\% & \underline{77.9\%} & 80.0\% & \underline{78.2\%} & \underline{73.2\%} & 84.4\% \\
Req. debt & \underline{12.0\%} & 14.9\% & \underline{1.0\%} & 5.9\% & \underline{1.1\%} & 4.5\% \\
Doc. debt & \underline{4.3\%} & 5.6\% & \underline{11.1\%} & 11.3\% & 19.3\% & \underline{7.5\%} \\
Test debt & 3.2\% & \underline{1.6\%} & 7.7\% & \underline{4.5\%} & 6.4\% & \underline{3.5\%} \\
\hline
\end{tabular}
}
\end{center}
\end{table}

We first compare the percentages of different types of SATD in industrial projects (IP) and open-source projects (OSP). The comparison is presented in \cref{tb:satd_perc_comparison}.
It is noted that data from industrial projects are calculated based on \cref{tb:num_of_td}, while the open-source data are obtained from 103 open-source projects from our previous work \cite{li2022automatic}.
Specifically, these 103 open-source projects are from the Apache ecosystem.
They are of high quality and well-maintained by mature communities.
Observing the table, we can find that the majority of SATD is \emph{code/design debt} in both industrial and open-source projects, ranging from 77.9\% to 84.4\% and 73.2\% to 80.6\% respectively.
Moreover, we notice that the second most prevalent types of SATD from different sources are consistent when comparing industrial and open-source projects.
Specifically, \emph{requirement debt} is the second most popular SATD type from code comments in both kinds of projects, while \emph{documentation debt} is the second most prevalent type of SATD from issues and commit messages in both kinds.

\begin{framed}
\noindent \textit{\textbf{Implication 1:}
The majority of SATD in both industrial and open-source projects is \textbf{code/design debt}.
The second most prevalent types of SATD from different sources (\textbf{requirement debt} or \textbf{documentation debt}) are also the same in the two settings.
Researchers could further investigate whether the types of SATD are similar in the two settings.
}
\end{framed}

Regarding differences between the two settings, in \cref{tb:satd_perc_comparison}, we observe the following: 1) the percentages of \emph{requirement debt} from different sources in industrial projects are significantly higher in comparison with open-source projects; 2) the percentages of \emph{test debt} of industrial projects are lower than open-source projects in the different sources.
For the differences in \emph{requirement debt}, we conjecture that this might result from the relatively high level of difficulty in changing embedded systems \cite{graaf2003embedded} and the high pressure of the studied projects as mentioned in \cref{sec:satd_attitude}: \textit{``I know [this] project is in challenging phase; they are high pressured to reach the time-to-market, [so] we are also under pressure to have shortcuts and do not redesigns [unless we are] told necessary by the developers.''}
Similarly, Zampetti \textit{et al.} \cite{zampetti2021self} found that industrial developers reported releasing software under more pressure compared to open-source developers. 

Moreover, the lack of self-admitting test debt in industry is consistent with our findings in \cref{sec:timepoint_doc_td} that \emph{certain types of technical debt are ignored} as some developers mentioned: \textit{``I think we don't have that many technical debt items for missing test cases; I think you more or less know about them, but no real documentation about test cases and actual implementation.''}
However, the reasons behind this phenomenon need further investigation.
Additionally, according to the differences in percentage between different types of SATD, as well as the interviews, we advise practitioners to create thresholds based on the percentages of different SATD types to evaluate the quality of SATD management.
For example, if there is significantly less test debt documented than the threshold and the code analysis tool shows low coverage, this might refer more to the reluctance of developers to admit test debt rather than low test debt.

\begin{framed}
\noindent \textit{\textbf{Implication 2:}
The percentage of \textbf{requirement debt} is higher while the percentage of \textbf{test debt} is lower in the studied industrial projects in comparison with open-source projects.
The differences between percentages of different types of SATD between different projects should be further studied to potentially create thresholds for evaluating the quality of SATD management.
}
\end{framed}

\begin{table}[th]
\caption{Comparison Between SATD Percentages in Different Sources in Industrial Projects (IP) and Open-Source Projects (OSP).}
\label{tb:percentage_comparison}
\begin{center}
\resizebox{\columnwidth}{!}{
\def\arraystretch{1.2}
\begin{tabular}{C{1.5cm}C{0.75cm}C{0.75cm}C{2.0cm}C{2.1cm}}
\hline
\multirow{2}{*}{Source} & \multicolumn{4}{c}{Percentage of SATD} \\
\cline{2-5}
 & OSP & IP & Percentage Diff. & Chi-Square Test \\
\hline
Comment & 5.2\% & \underline{2.6\%} & -50\% & $\chi^2(1)=2150.66$ $p<0.00001$ \\
Issue & \underline{13.0\%} & 16.3\% & +25.4\% & $\chi^2(1)=666.83$ $p<0.00001$ \\
Commit & \underline{11.3\%} & 12.7\% & +12.4\% & $\chi^2(1)=37.33$ $p<0.00001$ \\
\hline
\end{tabular}
}
\end{center}
\end{table}

Subsequently, we compare the percentages of SATD (irrespectively of type) in industrial projects and open-source projects, as shown in \cref{tb:percentage_comparison}.
Specifically, the data from industrial projects are obtained from \cref{tb:percentage_of_td}, while the data from open-source projects are still acquired from 103 open-source projects from our previous study \cite{li2022automatic}.
We perform chi-square tests to compare the number of SATD items from different sources in open-source and industrial projects.
As can be seen in \cref{tb:percentage_comparison}, the percentages of technical debt admitted in industrial and open-source projects are comparable.
Interestingly, the percentage of SATD from code comments in industrial projects is just half of the percentage of open-source projects (2.6\% versus 5.2\%); this is statistically significant (\textit{p}-value \textless~0.00001). Thus, less technical debt is admitted in code comments in industrial projects compared to open-source projects.
In a recent study, Zampetti \textit{et al.} \cite{zampetti2021self} investigated the preferences for documenting technical debt in source code comments between industrial developers and open-source developers.
They surveyed 101 software developers and found open-source developers tend to admit more technical debt in code comments in comparison to industrial developers.
Our results confirm these findings.

\begin{framed}
\noindent \textit{\textbf{Implication 3:}
The overall percentage of technical debt admitted in industrial and open-source projects are \textbf{comparable}.
There is, however, less technical debt admitted in \textbf{code comments} in industrial projects compared to open-source projects (2.6\% versus 5.2\%).
}
\end{framed}

Furthermore, we can see that significantly more technical debt is admitted in issues and commits (+25.4\% and +12.4\% respectively) in the industrial projects compared to the open-source ones in \cref{tb:percentage_comparison}.
This is likely related to the practices for SATD management that are followed in the two types of projects. 
Within our industrial partner, we established a preference for documenting SATD in issues for better tracking.
However, there might be other factors that have an impact on technical debt documentation.
Thus, the evidence shows that more technical debt is documented in issues and commits within the studied industrial projects compared to open source projects.
However, these differences are not as large as the differences in source code comments (25.4\% and 12.4\% versus 50\%).
Ultimately developers have the need to document SATD somewhere: industrial developers seem to prefer issues and commits while OS developers have a preference for source code comments. 
Researchers could further investigate if this is true more generally with further studies in industry.

\begin{framed}
\noindent \textit{\textbf{Implication 4:}
There is more technical debt admitted in \textbf{issues} and \textbf{commits} in industrial projects compared to open-source projects (+25.4\% and +12.4\%).
However, the differences are not as large as the differences in source code comments.
Researchers can investigate the differences in other projects and look deeper into the causes of these differences.
}
\end{framed}

Subsequently, we compare our results regarding the issue closing times and issue closing percentages with previous work.
There are two studies that investigated the closing time of issues.
The first one was conducted by Bellomo \textit{et al.} \cite{bellomo2016got}, who hypothesized that technical debt issues take a longer time to resolve than non-technical debt issues.
However, they found that the average open days of issues vary greatly, and their results did not support their hypothesis.
In the other study by Xavier \textit{et al.} \cite{xavier2020beyond}, they found that the median time to close technical debt issues is longer than other issues (16.7 days versus 4.0 days).
In this paper, we investigated the same research question in industrial settings (see \cref{sec:td_characteristics}).
The results shown in \cref{tb:time_and_pct_close,tb:time_to_close} indicate that technical debt issues take a longer time to resolve compared to non-technical debt issues (with statistical significance).
This supports the hypothesis proposed by the previous study \cite{bellomo2016got}.
It is noted that our study focuses on SATD in industrial settings.
This hypothesis still needs to be tested in open-source settings.
Furthermore, our study confirmed that the reason why technical debt issues take a longer time to close and have a lower closing rate is that developers are under pressure to implement new features or fix bugs instead of paying back technical debt (see \cref{sec:satd_attitude}).

\begin{framed}
\noindent \textit{\textbf{Implication 5:}
The hypothesis that technical debt issues take a longer time to resolve than non-technical debt issues is supported by our study in industrial settings.
Researchers could further examine this hypothesis in open-source settings and compare the results.}
\end{framed}

Moreover, our study investigated the time to close issues and the closing rate of issues with different types of SATD and non-SATD.
The results show that certain types of SATD take a significantly longer time to resolve than certain other types of SATD.
Specifically, observing \cref{tb:time_and_pct_close,tb:time_to_close}, we notice that \emph{requirement debt} and \emph{test debt} issues take a longer time to close compared to the other two types of SATD (70.2 and 80.7 days versus 62.5 and 60.4 days).
Moreover, they also have the first and second lowest closing rates compared to others (60.8\% and 67.0\% versus 71.3\% and 72.0\%).
Furthermore, we find that \emph{test debt} issues take a significantly longer time to resolve than \emph{code/design debt} and \emph{documentation debt} issues (\textit{p}-value is 0.014 and 0.020 respectively).
Overall, we observe that \emph{requirement debt} and \emph{test debt} might have a lower priority compared to \emph{code/design debt} and \emph{documentation debt}.
This finding is in agreement with the findings of Ebert and Jones \cite{ebert2009embedded} which showed that requirements and tests are the major cost drivers in embedded-software development (requirements engineering and testing take most of the effort).
As mentioned in \cref{sec:trigger_repay}, \emph{``repaying SATD takes too much effort''} is one of the triggers not to pay back SATD.
Thus, we conjecture that \emph{requirement debt} and \emph{test debt} issues take a longer time to close because they require more effort to repay than other types.
The detailed reasons behind the observations still need to be further investigated and validated.
Additionally, researchers could use our trained machine learning model to identify SATD issues in other projects and calculate the time spent to close different types of SATD issues, to investigate the priority of different types of SATD in different projects and compare results.

\begin{framed}
\noindent \textit{\textbf{Implication 6:}
\textbf{Requirement debt} and \textbf{test debt} take longer time to be solved compared to \textbf{code/design debt} and \textbf{documentation debt} in the studied industrial projects.
This observation still needs to be investigated and validated in other projects.
Researchers could also use our SATD analysis approach to study the priority of different types of SATD in open-source projects.
}
\end{framed}

In \cref{sec:satd_attitude}, we observed that most of the interview participants acknowledged the relevance and importance of the automatically-identified SATD items.
Because we utilized publicly available datasets \cite{da2017using,li2022identifying,li2022automatic} to train machine learning models to identify SATD from the industrial projects, the results show that our models can be used to accurately identify SATD in industrial projects.
We thus encourage researchers to use our approach to study SATD in industrial settings.
To facilitate this, we share our scripts and trained machine learning model in the replication package\cref{l:data}.
Moreover, we noticed that most of the participants (five out of nine) indicated that it is difficult to determine the importance of SATD based solely on SATD statements.
We suggest that researchers find the best method for presenting SATD, e.g. showing the SATD statements together with other contextual information to provide a broader picture.

\begin{framed}
\noindent \textit{\textbf{Implication 7:}
Most of the interviewees acknowledged the automatically identified SATD.
Researchers could use our trained machine learning model to further investigate SATD in other industrial projects.}
\end{framed}

In \cref{sec:relations}, we observed that it is common for source code comments or commit messages to reference related technical debt items in issues; the opposite is not common.
Moreover, developers believed it could be very useful if related SATD is linked together.
In the current state of the art, only the relation between code comments and commits has been used to study the repayment of SATD \cite{zampetti2018self,iammarino2019self,iammarino2021empirical}.
Thus, we argue that researchers and practitioners need to study the relations between SATD in different sources and build tools that aid in establishing traces between SATD in different sources and properly visualizing them.

\begin{framed}
\noindent \textit{\textbf{Implication 8:}
Researchers and practitioners could further investigate the relations between SATD in different sources and build tools to automate and visualize traceability between SATD in different sources.}
\end{framed}

\subsection{SATD Management Activities}

In \cref{sec:timepoint_doc_td} we saw seven distinct cases that cause technical debt to be ignored and stay undocumented. 
Such implicit technical debt can have grave consequences for the development team.
Researchers could look into this and propose solutions to avoid missing documentation for important technical debt.
Moreover, as pointed out in \cref{sec:challenges}, there are no concrete guidelines on technical debt documentation.
Our work extended the scope of SATD documentation to three sources, namely code comments, commit messages, and issue trackers.
Thus, we suggest that researchers and practitioners create comprehensive guidelines and develop tools to help technical debt documentation for different use cases in different sources based on our findings.
Furthermore, the pros and cons of documenting technical debt in different sources, as described in \cref{sec:pro_con}, can also be of assistance in creating the aforementioned guidelines and tools, by building on the pros and working to avoid the cons. 

\begin{framed}
\noindent \textit{\textbf{Implication 9:}
Researchers and practitioners could create guidelines and build tools to assist in technical debt documentation in different sources.}
\end{framed}

In \cref{sec:trigger_repay}, we reported the triggers for paying back and not paying back SATD.
However, the interviewees' opinions on \emph{certain types of SATD} are contradictory.
More specifically, some developers saw \emph{test debt} as a trigger to repay SATD, while some others believed it should be ignored.
This is caused by different participants' opinions about the importance of \emph{test debt}.
We encourage researchers and practitioners to create guidelines customized for specific organizations and teams about the importance and ways of repaying different types of SATD.
The triggers identified in our study can act as an input for such guidelines.
Moreover, researchers could also study how to eliminate the effects of SATD accumulation caused by triggers for not paying back technical debt.
Finally, researchers could investigate the triggers in open-source projects, compare them, and create comprehensive lists of triggers for both open-source and industrial projects.

\begin{framed}
\noindent \textit{\textbf{Implication 10:}
Researchers could investigate triggers for repaying and not repaying SATD, and create guidelines and tools for SATD prioritization based on those triggers.
Besides, strategies to mitigate the effects of SATD accumulation caused by triggers for not paying back SATD need to be studied.
Moreover, researchers could investigate triggers in open-source projects and compare results.
}
\end{framed}

In \cref{sec:satd_management}, we reported the practices used to support SATD prioritization.
These practices are not acknowledged by all the interviewees.; for example, in contrast to the practice of using issue types (e.g., \emph{bug}, \emph{task}, and \emph{backlog item}) to set priorities of SATD, one participant did not find significant differences between issue types. This is due to the organization not using such issue types in a standardized manner. 
Researchers should study and propose such practices for SATD prioritization that can be standardized across organizations.

\begin{framed}
\noindent \textit{\textbf{Implication 11:}
Researchers could study and use the practices to support SATD prioritization by embedding them in tools or processes.}
\end{framed}

In \cref{sec:satd_management}, we also report on two strategies for efficiently paying back SATD.
We found that adopting such strategies heavily depends on developers' personal opinions and discussions with their colleagues.
Researchers could investigate how much effort (i.e. technical debt interest) is saved by using these strategies and how to automatically group SATD and other tasks for higher repayment efficiency.

\begin{framed}
\noindent \textit{\textbf{Implication 12:}
Researchers could investigate the efficiency of SATD repayment strategies and build tools to help developers utilize these strategies.}
\end{framed}

\subsection{SATD Management Improvement}

In \cref{sec:challenges}, we list the challenges faced by interviewees when dealing with SATD.
We suggest that researchers examine the impact of the listed challenges, and propose strategies and tools to tackle them. Some of the challenges can be addressed directly within the development team, e.g. convincing developers not to introduce technical debt when not necessary. Practitioners could discuss these challenges in their team and decide which they can tackle and how.

\begin{framed}
\noindent \textit{\textbf{Implication 13:}
Researchers can evaluate the impact of challenges, and propose strategies and tools to tackle them. Practitioners can also review them and discuss which ones they can address.}
\end{framed}

\begin{table}[t]
\caption{Comparison Between Suggested Features and State-Of-The-Art.}
\label{tb:work_comparison}
\begin{center}
\resizebox{\columnwidth}{!}{
\def\arraystretch{1.5}
\begin{tabular}{R{2.6cm}L{3.5cm}L{3.0cm}}
\hline
\multicolumn{2}{c}{Suggested Features} & Related Papers \\
\hline
\multirow{4}{*}{\makecell[r]{Automated SATD\\Identification From}} & Code Comments & \cite{da2017using,freitas2016investigating,huang2018identifying,wattanakriengkrai2018identifying,liu2018satd,de2018study,yan2018automating,ren2019neural,wattanakriengkrai2019automatic,flisar2019identification,guo2019mat,wang2020detecting,rantala2020predicting,alhefdhi2020framework,de2020identifying,maipradit2020wait,maipradit2020automated,santos2020long,yu2020identifying,rantala2020prevalence,chen2021multiclass,guo2021far,santos2021evaluating,yu2021using,sala2021debthunter,zhu2021detecting,phaithoon2021fixme,xiao2021characterizing,yu2022exploiting,li2022automatic,tu2022debtfree,russo2022weaksatd,alomar2022satdbailiff,zhuang2022empirical} \\
\cline{2-3}
 & Issue Trackers & \cite{dai2017detecting,li2022identifying,li2022automatic} \\
\cline{2-3}
 & Commit Messages & \cite{li2022automatic} \\
\cline{2-3}
 & Pull Requests & \cite{li2022automatic} \\
\hline
\multicolumn{2}{r}{\makecell[r]{Automated Differentiation Between Fixed and \\Unfixed SATD}} & - \\
\hline
\multirow{3}{*}{\makecell[r]{Automated Tracing\\Between SATD}} & in Different Sources & \cite{zampetti2018self,iammarino2019self,iammarino2021empirical,li2022automatic} \\
\cline{2-3}
 & and Code & - \\
\cline{2-3}
 & and Related Development Tasks & - \\
 \hline
\multicolumn{2}{r}{Automated SATD Prioritization} & \cite{zampetti2017recommending,mensah2018value,de2022toward} \\
 \hline
\multicolumn{2}{r}{Automated Identification of SATD Risk} & - \\
 \hline
\multicolumn{2}{r}{Automated Estimation of Benefits to Solve SATD} & \cite{kamei2016using} \\
 \hline
\multicolumn{2}{r}{Automated Estimation of Cost to Solve SATD} & \cite{mensah2018value,mensah2016rework} \\
 \hline
\multicolumn{2}{r}{Automated SATD Solution Suggestion} & \cite{zampetti2020automatically} \\
\hline
\end{tabular}
}
\end{center}
\end{table}

The participating developers have come up with various features they would like to see in SATD management tools (see \cref{sec:tools}).
This begs the question of whether the current research work can already offer some of these features. To offer a preliminary answer, we checked research publications in Google Scholar, using the search string \emph{``self-admitted technical debt''}.
This resulted in 72 papers that deal with SATD; we then read their abstract and full text to filter out papers that are irrelevant to the required features (see \cref{sec:tools}).
The resulting set of 45 related papers is listed in \cref{tb:work_comparison}.
As we can see, the majority of these papers focus on automatically identifying SATD from source code comments, while there has been no work investigating \emph{automated differentiation between fixed and unfixed SATD}, \emph{automated tracing between SATD and code or related development tasks}, and \emph{automated identification of SATD risk}.
For the other proposed features, some related studies exist, but these are not fully supported yet or require better support.
For example, there is a study relevant to \emph{automated SATD solution suggestion} \cite{zampetti2020automatically}, but it is only able to suggest one of six predefined SATD repayment strategies (e.g., changing API calls or changing return statements).
Furthermore, the usefulness of each feature and the difficulties of implementing each feature are different.
We recommend that researchers and practitioners evaluate the added value of each proposed feature, implement tools including the most important features, and test the effectiveness of such tools.

\begin{framed}
\noindent \textit{\textbf{Implication 14:}
Most research works in SATD management tools focus on automatically identifying SATD from source code comments.
Researchers should investigate other features required by the interviewees, such as automated differentiation between fixed and unfixed SATD, automated tracing between SATD and code or related development tasks, and automated identification of SATD risk.
}
\end{framed}

\section{Threats to Validity}
\label{sec:validity}

\subsection{Construct Validity}

Threats to construct validity concern the correctness of operational measures for the studied subjects.
One of the threats to construct validity in the study concerns the potentially different interpretations of discussed topics between interviewees and researchers.
Because we focus on SATD in this study and most of the interviewees were not familiar with this concept, we tried to avoid misunderstanding this term by 1) asking for their understanding of technical debt; 2) asking them to give some examples of it to make sure they have the correct comprehension; 3) reminding them, during the interviews, that we focus on technical debt that is documented in different sources to avoid confusion.
The responses we received from the interviewees regarding how they understand technical debt and the examples of technical debt they gave, attest to a correct understanding of the concept by all interviewees. For instance, one of the interviewees gave the following examples of technical debt: \textit{``for me, technical debt can be multiple things; it can be a design that works now but might be problematic in the future. It also covers shortcuts you take in the code. When you have a bug, you create a quick workaround, so your customers can continue, which you know actually [...] needs a solid fix so that in the future it will remain intact. I think technical debt also builds automatically if certain techniques are no longer maintained, so you need to switch to a new one because you cannot upgrade it...''}.

Another threat to construct validity is related to the possible reluctance of interviewees to express negative opinions on their organization or admit mistakes made in the past.
To minimize this threat, we emphasized that we are bound by a confidentiality agreement, and no sensitive or personal information would be revealed after the interviews.

\subsection{Reliability}

Reliability is concerned with the bias that researchers may induce in data collection and analysis.
One threat to reliability could be different results obtained from work artifacts' analysis.
Specifically, when extracting source code comments from studied projects, because the projects could use multiple programming languages, defining and using simple heuristic rules might not be able to extract all comments from different programming languages.
To reliably and accurately extract code comments, instead of defining such heuristic rules ourselves, we chose to use a third-party library (\emph{CommnetParser}), which supports multiple programming languages, such as C++, Go, Python, Java, XML, and Ruby.
The studied projects mainly use C++ and XML files.
We manually verified the correctness of the extracted comments with this library before collecting the data.

Another threat to reliability could be the impact of researchers' opinions on interviewees.
To mitigate this threat, all authors followed a specific protocol for the interviews which is included in the replication package\cref{l:data}. 
Besides, at least two authors attended each interview, to ensure that one interviewer did not bias the questions asked.

Furthermore, another threat to reliability could come from the selection of the 15 SATD items for interviews.
As we can see in \cref{tb:satd_perc_comparison}, the percentages of requirement, documentation, and test debt are relatively low (always below 15\%).
If we randomly collected five SATD items from each source, certain types of SATD items would likely be missing in the 15 SATD samples.
This could result in the sample misrepresenting the SATD types.
To mitigate this risk, we selected three or four SATD items for code/design debt and one or two SATD items for other types of SATD for each source (e.g., code comments) based on the SATD type proportion (see \cref{tb:satd_perc_comparison}).
Thus, for the 15 SATD sample items, we have ten code/design debt items, one requirement debt item, two documentation debt items, and two test debt items, which include different types of SATD items and follow the distribution of different types of SATD. Therefore, we consider this threat as, at least partially, mitigated.

The last threat to reliability comes from analyzing the interview data.
To minimize this threat, the first and second authors carried out the \emph{Constant Comparative} analysis process \cite{strauss1990basics,stol2016grounded} independently; in case of discrepancy, we compared and discussed the results until we were able to reach an agreement.

\subsection{External Validity}

Threats to external validity concern the generalizability of the results.
In this study, we analyzed work artifacts and conducted interviews in a large software company in the embedded systems industry.
Our findings may, to some extent, generalize to other industrial projects of this application domain and of similar size and complexity.
In several instances, such as time to close SATD and non-SATD issues and the percentage of SATD in code comments, we have demonstrated how our findings support previous studies.
However, we can not claim any further generalization.

\section{Conclusions and Future Work}
\label{sec:conclusion}

In this paper, we analyzed SATD in industrial projects using machine learning techniques and conducted 12 interviews to understand: 1) characteristics of SATD in industrial projects; 2) developers' attitudes towards identified SATD and statistics; 3) triggers to introduce and repay SATD; 4) relations between SATD in source code comments, issues, and commits; 5) practices used to manage SATD; 6) challenges and tooling ideas for SATD management.

The results present characteristics of SATD in industrial projects and shed light on developers' opinions on SATD management and tooling support.
This promotes future studies in this area, targeting to support developers in terms of SATD introduction, traceability, prioritization, repayment, and tool support.

In the future, based on the results of this work, we plan to characterize SATD in more industrial projects to further validate the observation that industrial developers tend to admit less SATD in code comments and more SATD in issues and commits in comparison with open-source developers, and explore the reasons behind it.
Moreover, we plan to conduct a large-scale study to analyze the closing time and closing rate between different types of SATD in open-source projects, in order to investigate the priority differences between different types of SATD and non-SATD issues.
Furthermore, we plan to study the relations between SATD in different sources and create an automatic approach to identify related SATD items.
Additionally, we plan to create SATD documentation and repayment guidelines and evaluate them with software practitioners.
Next, we plan to investigate the number of SATD items that are documented or repaid in terms of the different reasons to document SATD or pay back SATD.
Lastly, we plan to build a tool that supports SATD management for software practitioners based on the desired features described in \cref{sec:tools}.

\bibliography{bibliography}
\bibliographystyle{IEEEtran}

\end{document}